\documentclass[journal=jpcbfk,manuscript=article]{achemso}
\usepackage{textcomp}
\usepackage{longtable}
\usepackage{multirow}
\usepackage{amsmath}
\usepackage{subfigure}
\usepackage[usenames,dvipsnames]{color}
\usepackage{rotating}
\usepackage{threeparttable}
\usepackage{xr}

\SectionNumbersOn

\author{Silvan K\"aser} \affiliation[University of Basel]{Department
  of Chemistry, University of Basel, Klingelbergstrasse 80 , CH-4056
  Basel, Switzerland.}

\author{Oliver T. Unke} \affiliation[University of Basel]{Department
  of Chemistry, University of Basel, Klingelbergstrasse 80, CH-4056
  Basel, Switzerland\\ Present Address: Machine Learning Group, TU
  Berlin, Marchstr. 23, 10587 Berlin, Germany}

\author{Markus Meuwly} \affiliation[University of Basel]{Department of
  Chemistry, University of Basel, Klingelbergstrasse 80 , CH-4056
  Basel, Switzerland.}  \email{m.meuwly@unibas.ch}

\title {Isomerization and Decomposition Reactions of Acetaldehyde
  Relevant to Atmospheric Processes from Dynamics Simulations on
  Neural Network-Based Potential Energy Surfaces}

\begin{document}

\date{\today}

\begin{abstract}
Acetaldehyde (AA) isomerization (to vinylalcohol, VA) and
decomposition (into either CO+CH$_4$ and H$_2$+H$_2$CCO) is studied
using a fully dimensional, reactive potential energy surface
represented as a neural network (NN). The NN, trained on 432'399
reference structures from MP2/aug-cc-pVTZ calculations has a MAE of
0.0453~kcal/mol and an RMSE of 1.186~kcal/mol for a test set of 27'399
structures. For the isomerization process AA$\rightarrow$VA the
minimum dynamical path implies that the C–H vibration, and the C–C–H
(with H being the transferring H-atom) and the C–C–O angles are
involved to surmount the 68.2 kcal/mol barrier. Using an excess energy
of 93.6 kcal/mol -- the energy available in the solar spectrum and
sufficient to excite to the first electronically excited state -- to
initialize the molecular dynamics, no isomerization to VA is observed
on the 500 ns time scale. Only with excess energies of $\sim 127.6$
kcal/mol (including the zero point energy of the AA molecule),
isomerization occurs on the nanosecond time scale. Given that
collisional de-excitation at atmospheric conditions in the
stratosphere occurs on the 100~ns time scale, it is concluded that
formation of VA following photoexcitation of AA from actinic photons
is unlikely. This also limits the relevance of this reaction pathway
to be a source for formic acid.
\end{abstract}

\section{Introduction}
Understanding and quantifying the relative importance of
tautomerization versus decomposition reactions of small organic
molecules is an essential aspect in modeling atmospheric
processes\cite{vereecken2018perspective}. Recently, the
photo-tautomerization of acetaldehyde (AA) to vinyl alcohol (VA) has
been investigated experimentally\cite{shaw2018photo} and the results
were linked to formic acid (FA) production in the
atmosphere.\cite{shaw2018photo,archibald:2007} Although the
experiments provided insight into the tautomerization dynamics, the
time scale on which the AA$\rightarrow$VA isomerization occurs is
still unknown.\\

\noindent
In general, the role of photo-induced tautomerization of carbonyls in
atmospheric modeling is incompletely
characterized.\cite{kable:2012,osborn:2012} In such systems,
interconversion barrier heights for tautomerization and fragmentation
can be similar which leads to competing reaction pathways. This
situation is similar to halogenated sulfur-containing compounds (such
as H$_2$SO$_4$ or HSO$_3$F) for which proton transfer barrier heights
(isomerization) only differ by a few kcal/mol compared with those
leading to SO$_3$+H$_2$O (for H$_2$SO$_4$) or HF+H$_2$O (for
HSO$_3$F),\cite{Miller:2006,nagy2014multisurface,MM:2014.h2so4,reyes2016hso}
important processes that have also been implicated in atmospheric
chemistry.\cite{vaida.sci.2003.vibphotodis} Such a topography of the
potential energy surface (PES) leads to rich molecular dynamics and a
broad distribution of reaction times for the fragmentation process due
to internal vibrational energy redistribution (IVR).\\

\noindent
For a molecularly refined understanding of the reaction dynamics and
for determining the corresponding rates, accurate and
fully-dimensional PESs for reactive atomistic simulations are
required. An essential prerequisite for molecular dynamics (MD)
simulations is the availability of energies and corresponding forces
for an extensive range of molecular configurations, which can either
be obtained from solving the electronic Schr\"odinger equation at
every time step of an MD simulation (QM/MD or ab initio MD) or by
parametrizing a suitable reactive force field such as in multi state
adiabatic reactive MD.\cite{nagy2014multisurface}\\

\noindent
Due to the unfavourable scaling of high-accuracy electronic structure
methods with the number of electrons (and basis functions required for
an accurate description) it is only possible to use {\it ab initio} MD
methods for small molecules in the gas phase and for short time scales
(several 10 picoseconds). \cite{francisco:2019} These limitations
often make examining interesting chemical or physical problems
difficult. To overcome this it is necessary to find computationally
efficient representations of the intermolecular interactions to
describe the energies of molecular configurations. One such approach
are reactive MD simulations\cite{revrmd.2012,meuwly2019reactive} which
allow to study the breaking and formation of chemical bonds.\\

\noindent
More recently, artificial neural networks (ANNs) have emerged as an
alternative way to fit and calculate the energies of molecules based
on a large number of reference electronic structure calculations. NNs
were introduced more than half a century
ago\cite{mcculloch1943logical,rosenblatt1958perceptron} and are a set
of biology-inspired algorithms which learn patterns and interrelations
from extensive data.\cite{behler2011neural,schalkoff1997artificial}
Because they are general function
approximators\cite{hornik1989multilayer} they constitute an ideal tool
for generating representations of molecular PESs.\\

\begin{figure}[h!]
\centering \includegraphics[width=1\textwidth]{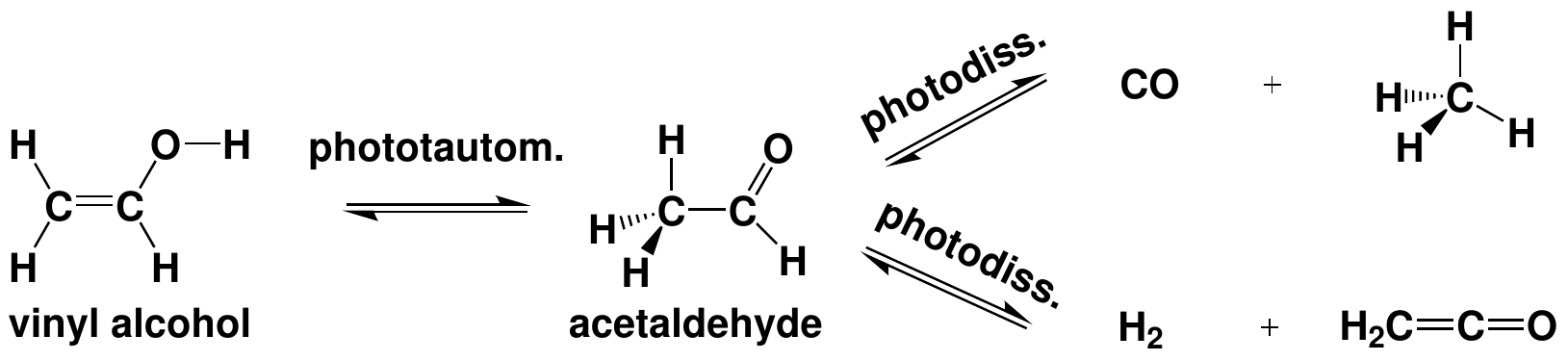}
\caption{Schematic molecular representation of the tautomerization and
  dissociations of acetaldeyhde (AA, middle). Here, isomerization to
  vinylalcohol (VA, left) and dissociation to CO+CH$_4$ and
  H$_2$+C$_2$OH$_2$ (right) is studied}
\label{fig:acetreac}
\end{figure}

\noindent
In the present work a reactive, global NN-based PES is developed for
acetaldehyde (AA) to study tautomerization and decomposition
reactions. AA (CH$_3$CHO) plays an important role in atmospheric
chemistry by adversely affecting global
climate.\cite{finlayson1999chemistry} The smallest organic acids -- FA
and acetic acid -- make up a significant fraction of the tropospheric
organic acids. The majority of atmospheric organic acids are believed
to be generated via photochemical oxidation of biogenic and
anthropogenic volatile organic compounds (VOCs). However, the global
budget for FA indicates that in addition to the largest source for FA
(e.g.\ photochemical production through fires, oxidation of organic
aerosols) other formation pathways must contribute.\cite{millet:2015}
One proposed formation pathway is the generation of FA via oxidation
by the hydroxyl radical\cite{shaw2018photo,dasilva:2014} following
photo-tautomerization of AA to its enol form
VA.\cite{archibald:2007,kable:2012,osborn:2012} The relevant states
for tautomerization of AA to VA, the dissociation to carbon monoxide
and methane, and the dissociation to H$_2$ and ketene are shown in
Figure \ref{fig:acetreac}.\\

\noindent
Recent laboratory experiments under atmospheric conditions reported
direct evidence that AA does photo-tautomerize to VA. To this end,
pressure- and wavelength-dependent quantum yields for VA production in
up to 1 atm of N$_2$ in the actinic wavelength range of $\lambda =
300$ to 330 nm (corresponding to an excitation energy of $\sim 95$
kcal/mol) were determined.\cite{shaw2018photo} One of the pertinent
questions is therefore whether isomerization from an electronically
excited AA to VA competes kinetically with collisional de-excitation
through other molecules in the atmosphere and other competing
intramolecular pathways such as decomposition into CO and CH$_4$ or
into H$_2$ and H$_2$CCO. For this, typical isomerization times are
required which have so far not been obtained from experiment but can
be determined from reactive MD simulations. To carry out such
simulations, a reliable, fully dimensional reactive PES is required in
order to follow the nonequilibrium dynamics after electronic
excitation and relaxation of AA into relevant states -- including VA,
CH$_4$+CO, and H$_2$+H$_2$CCO, see Figure \ref{fig:acetreac}.\\

\noindent
Previous computational studies using QCT simulations focused on the
final state analysis\cite{bowman:2007} and branching
ratio\cite{bowman:2012} at high energies for several reaction
products. The simulations were based on a PES fitted to
permutationally invariant polynomials.\cite{bowman.irpc:2009} Other
related processes that were recently studied included formic
acid-assisted conversion of VA to AA\cite{peeters:2015} and the
oxidation of VA by the hydroxyl radical (OH) which was found to be
fast and generates mainly FA.\cite{dasilva:2014}\\

\noindent
In the present work a neural network-based PES is employed to carry
out reactive molecular dynamics simulations for a large number of
diverse initial conditions. In the following, first the construction
of the NN PES is described. Then, the minimum dynamical path
(MDP)\cite{unke2019sampling} connecting different states is determined
and reactive MDs are run and analyzed. Finally, the results are
discussed in the context of atmospheric chemistry.\\

\section{Methods}
First, the development of the NN-based, reactive PES for AA and its
isomerization (VA) and several reaction products (CH$_4$+CO,
H$_2$+H$_2$CCO) is described. This includes the generation of
reference data, as well as the architecture and training of the
NN. Section~2.2 outlines the
protocol used for the molecular dynamics (MD) simulations and their
analysis. All programs used for the reference data set and MD
simulations were written in the Python programming language and used
the Atomic Simulation Environment (ASE) \cite{larsen2017atomic}.\\

\subsection{Construction of the Potential Energy Surface}
\label{sec:construction_of_the_pes}
The present work uses PhysNet\cite{unke2019physnet}, a
high-dimensional NN\cite{behler2007generalized} designed to learn
molecular properties including energy, forces, and dipole
moments from \textit{ab initio} reference data, to construct the
PES. In the following, its architecture and the generation of
reference data are briefly summarized. For more details the reader is
referred to Reference~\citenum{unke2019physnet}. \\

\noindent
PhysNet predicts atomic energy contributions and partial charges based
on feature vectors that encode the local chemical
environment\cite{unke2018reactive} of each atom $i$. These features
are constructed by ``passing messages''\cite{gilmer2017message}
between all atoms within a cutoff radius of $r_{\rm cut} = 10$~\r{A}
and encode information about nuclear charges $Z_i$ and Cartesian
coordinates $\mathbf{r}_i$. The total potential energy $E$ of the
system is the sum of all $N$ atomic contributions and includes
long-range electrostatics and dispersion interactions explicitly:
\begin{equation}
E =\sum_{i=1}^N E_i + k_e\sum_{i=1}^N\sum_{j>i}^N \frac{q_i
  q_j}{r_{ij}}+E_{\rm D3}
\label{eq:physnet_total_energy}
\end{equation}
Here, $E_i$ and $q_i$ are atomic energy contributions and partial
charges (corrected to guarantee charge conservation, see
Reference~\citenum{unke2019physnet}), $k_e$ is Coulomb's constant,
$r_{ij}$ is the distance between atoms $i$ and $j$ and $E_{\rm D3}$ is
Grimme's D3 dispersion correction.\cite{grimme2010consistent} To
prevent potential numerical instabilities due to the singularity at
$r_{ij} = 0$, the Coulomb term is damped for small distances (see
Reference~\citenum{unke2019physnet}, not shown in
Eq.~\ref{eq:physnet_total_energy}). The parameters for the dispersion
correction, which are adapted during the training procedure, are
initialized to the standard values recommended for the Hartree-Fock
level of theory.\cite{grimme2011effect} Since the atomic features
(from which $E_i$ and $q_i$ are derived) are constructed based only on
pairwise distances, and the influence of all atoms within the cutoff
is combined by summation, the energy $E$ is invariant with respect to
translation, rotation, as well as permutations of equivalent
atoms. Analytical derivatives of $E$ with respect to the Cartesian
coordinates $\mathbf{r}_i$, required for the forces in the MD
simulations, are obtained by reverse mode automatic
differentiation\cite{baydin2018automatic}.\\

\noindent
In order to construct a PES, \textit{ab initio} reference data is
required for training the NN. Here, the MP2/aug-cc-pVTZ level of
theory\cite{moller1934note,dunning1989gaussian} is used as a
compromise between speed and accuracy. Although single point
calculations would also be possible at a higher level of theory, such
as CCSD(T),\cite{bowman:2007} a lower level is preferred here because
the number of reference calculations required to reliably represent a
high-dimensional, reactive PES is unknown \textit{a priori}. As is
described further below, the final data set required to run stable MD
simulations contains more than 400'000 energies for the present work
which would be computationally very demanding to be determined at a
higher level. The reference data (energies, forces and dipole moments)
are calculated using the MOLPRO software
package.\cite{schuetz2018MOLPRO}\\

\noindent
Motivated by the ``amons'' approach \cite{huang2017dna}, a set of
molecules covering a range of fragmentation products and stable
intermediates is chosen (see Figure~S1). Then,
MD simulations for all these molecules and the van der Waals complexes
of the fragmentation products CH$_4$+CO and H$_2$+H$_2$CCO are run to
obtain a broad range of molecular geometries. All simulations start at
the optimized geometries and are propagated using Langevin dynamics at
1000~K using a time step of 0.1~fs. The forces are obtained at the
semi-empirical PM7\cite{stewart2007optimization} level of theory using
MOPAC\cite{stewar2016mopac}. For the van der Waals complexes, complete
dissociation is prevented using a harmonic potential, akin to umbrella
sampling.\cite{torrie1977nonphysical} In the same way, structures
around the transition states (TSs) are sampled by harmonically biasing
the geometry towards the TS structure.\\

\noindent
On this initial data set, two independent NNs are trained and the MD
simulations are repeated using their averaged predictions to obtain
the necessary forces. Then, the data set is extended based on adaptive
sampling\cite{behler2016perspective,behler2015constructing}: If the
energy predictions of both NNs differ by more than a threshold value
(here $0.5$~kcal/mol) during the simulations, this indicates that the
reference data is insufficient to describe this part of the PES. The
corresponding structures are saved and new \textit{ab initio}
calculations are performed and added to the data set. Then, the NNs
are re-trained on the extended data set and the sampling is repeated
until differences (0.5 kcal/mol, see above) between the NN predictions
are either rare or nonexistent. After the third iteration, the dataset
is cleaned by removing outliers from the ensemble of networks for
which $|E_{\rm MP2}-E_{\rm NN}| > 3$ kcal/mol. Such structures with
large prediction errors are typically generated at an early stage of
the adaptive sampling when the global PES is still insufficiently
sampled. The last two iterations of structure sampling were performed
based on the normal mode sampling method.\cite{smith2017ani}\\

\noindent
After a total of six iterations, the final data set contained 432'399
structures. The data set was then split into a training (380'000), a
validation (25'000), and a test set (27'399) and used to train the
final NNs. The subsequent analysis and MD simulations are performed
using one NN, unless stated otherwise.\\

\subsection{Molecular Dynamics Simulations}
\label{sec:molecular_dynamics_simulations}
The final NN PES was first used to examine the minimum dynamical path
(MDP) and to run different types of MD simulations starting from a
range of initial conditions. The MDP corresponds to the path a
trajectory follows when going from a reactant to a product (or vice
versa) and passing through the exact transition state with zero excess
energy. \cite{unke2019sampling} It can be calculated by assigning
infinitesimal momenta along the normal mode vector with imaginary
frequency and then advancing the MD simulation from a TS. The MDP is
examined in order to characterize the internal degrees of freedom that
are dynamically coupled when transitioning from reactant to
product. For example, by analysing the MDP for a Diels-Alder reaction,
it was found - and subsequently confirmed from explicit MD simulations
- that relative rotational motion of the reactants promotes product
formation.\cite{rivero2019reactive} Such insights are difficult to
obtain from a static picture of a reactive process, e.g.\ via the
minimum energy path (MEP). \\

\noindent
All simulations were carried out in the microcanonical ($NVE$)
ensemble using the Velocity Verlet integrator\cite{verlet1967computer}
with a time step of $\Delta t = 0.1$~fs.  The first set of MD
simulations is started from the optimized AA structure. Random momenta
are assigned and scaled such that the total kinetic energy is $E_{\rm
  ex} = 93.6$ kcal/mol. This energy corresponds to the total energy
content of the molecule after returning to the ground state PES
following photoexcitation and is slightly lower than the maximum
excitation energy used in the experiments which is $\sim 95$
kcal/mol.\cite{shaw2018photo} 2000 trajectories with different initial
momenta are run for 250 ps each, yielding a total of 500 ns simulation
time. In the following, this set of simulations is referred to as
\textit{EX} (excitation) trajectories.\\

\noindent
For the second set of simulations, the energy of the system was
increased to 127.6 kcal/mol, as was done in
Reference~\citenum{shepler2007quasiclassical}. This energy includes
the excitation energy $E_{\rm ex}$ and the harmonic zero point energy
(ZPE) of AA. The ZPE of AA was also calculated for the NN PES
(35.1~kcal/mol) and found to be within 0.3~kcal/mol of the energy used
in Reference~\citenum{shepler2007quasiclassical} (34.8~kcal/mol at the
CCSD(T) level of theory). A total of 10'000 trajectories, each 50~ps
in length, are run, yielding a total simulation time of 500~ns. This
set of trajectories will be referred to as \textit{ZPE}.\\

\noindent
Because the MD simulations are run at large excess energy on the
ground state PES, the validity of the NN for the structures sampled
needs to be checked thoroughly. For this, all trajectories are
analyzed for unusual molecular structures -- such as unusually short
atom-atom separations or dissociations -- based on geometric
considerations. Moreover, the energy of single trajectories is
evaluated with two independently trained NNs and their energy
prediction is compared (see Figure~S2).\\

\section{Results}
First, the accuracy of the NN-PES is investigated. Next, the minimum
dynamical paths (MDPs)\cite{unke2019sampling} along the reaction
pathways for the tautomerization between AA and VA, the dissociation
to methane and carbon monoxide, and the dissociation to H$_2$ and
ketene (see Figure \ref{fig:acetreac}) are explored in order to
characterize the internal degrees of freedom that are involved in the
different atomic rearrangements. Finally, the two sets of trajectories
are analyzed and the likelihood for reaction is assessed.\\

\subsection{Quality of the Potential Energy Surface}
\label{sec:quality_of_the_pes}
\begin{figure}[h!]
\centering
\includegraphics[width=0.9\textwidth]{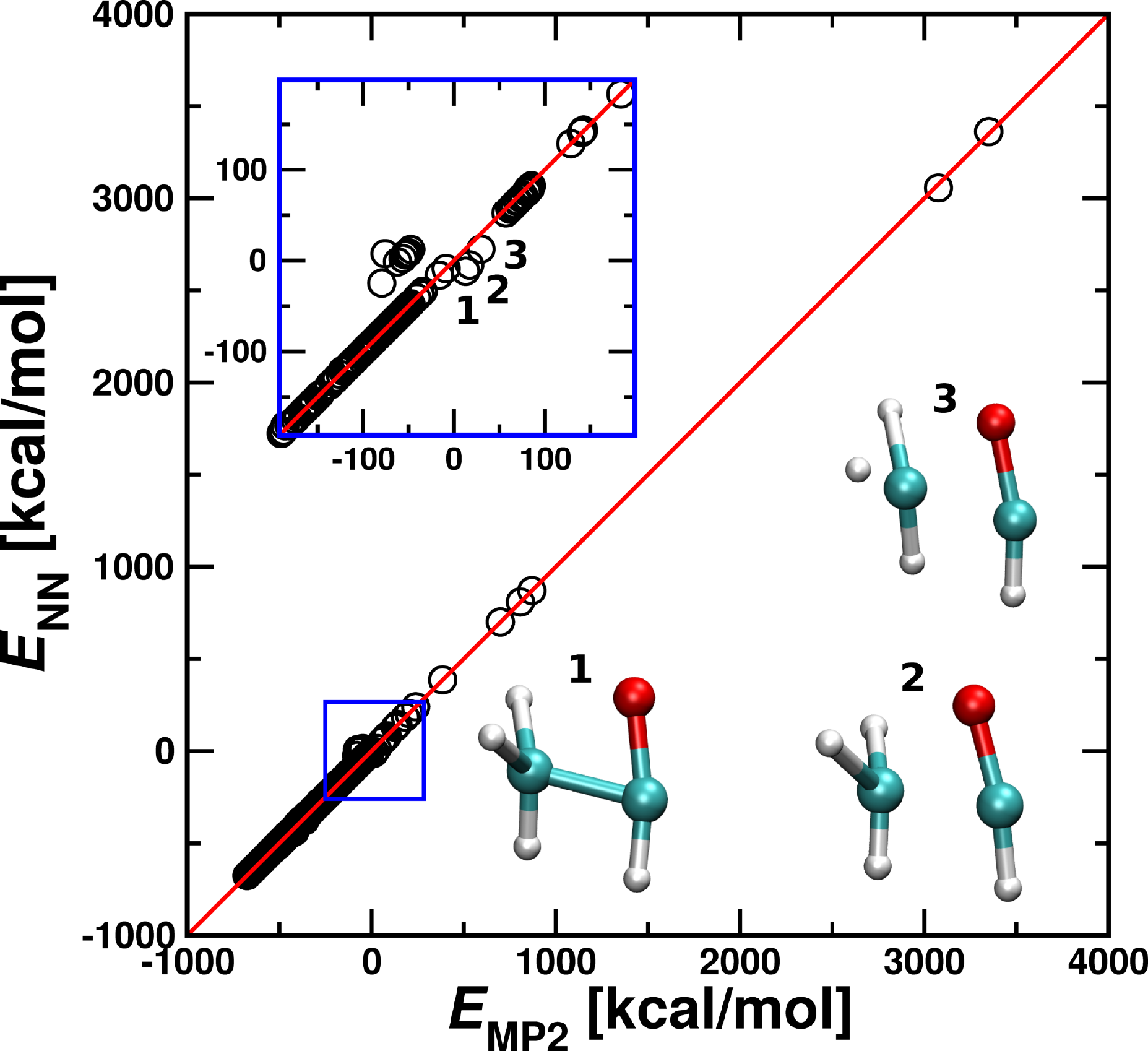}
\caption{Comparison of the MP2 reference and predicted NN energies for
  the test set. The 27399 structures from the test set are predicted
  with a MAE of 0.0453 kcal/mol, an RMSE of 1.186 kcal/mol and a
  Pearson coefficient of $1-2.86 \cdot 10^{-5}$. The graph is
  complemented with a zoom-in (blue) and three outlier structures (1
  to 3) and nine outliers above the diagonal. All outliers including
  their $T_1$ value\cite{lee1989diagnostic} are reported in
  Figure~S3. For structures in the insets,
  ``bonds'' are shown up to a distance of 2.6 \AA\/.}
\label{fig:corracetaldehyde}
\end{figure}

The accuracy of the NN PES is examined by comparing the predicted NN
energies with the reference MP2 energies for the 27'399 test set
structures, see Figure~\ref{fig:corracetaldehyde}. Most of the NN
energies match the MP2 energies, except for a few outliers. The MAE is
0.0453 kcal/mol with an RMSE of 1.186 kcal/mol and a Pearson
coefficient of $1-2.86 \cdot 10^{-5}$ when compared with the reference
calculations. Three outliers (below the red line) with unusually large
CH (up to 2.65 \AA\/), CO (up to 2.34 \AA\/), and CC (up to 2.72
\AA\/) separations are present in the test set. These were sampled in
the early adaptive sampling runs.\\

\noindent
Moreover, the enlargement reveals a group of nine structures with a
comparable error of $\approx 60$ kcal/mol (above red line). Sampling
structures far from equilibrium for the molecules considered
potentially leads into regions of the PES where multireference effects
can become relevant. To test this, the $T_1$
diagnostic\cite{lee1989diagnostic} is determined for these outliers
which are characterized by unusual geometries with elongated bonds
(see Figure~S3). A value for $T_1 > 0.02$
indicates that a single-reference wavefunction may be insufficient to
describe the system. The $T_1$ values were calculated at the (pair
natural orbital - local CCSD(T)) PNO-LCCSD(T)-F12/cc-pVTZ-F12 level of
theory.\cite{werner2018coupled-cluster} For most structures, the $T_1$
diagnostic is well above a value of 0.02
(Figure~S3). Hence, their multireference character
may be the reason for the large prediction errors. It has already been
shown earlier that NNs can be effective to identify members in a test
set that are likely outliers.\cite{unke2018reactive} \\

\begin{figure}[!ht]
\centering
\includegraphics[width=1\textwidth]{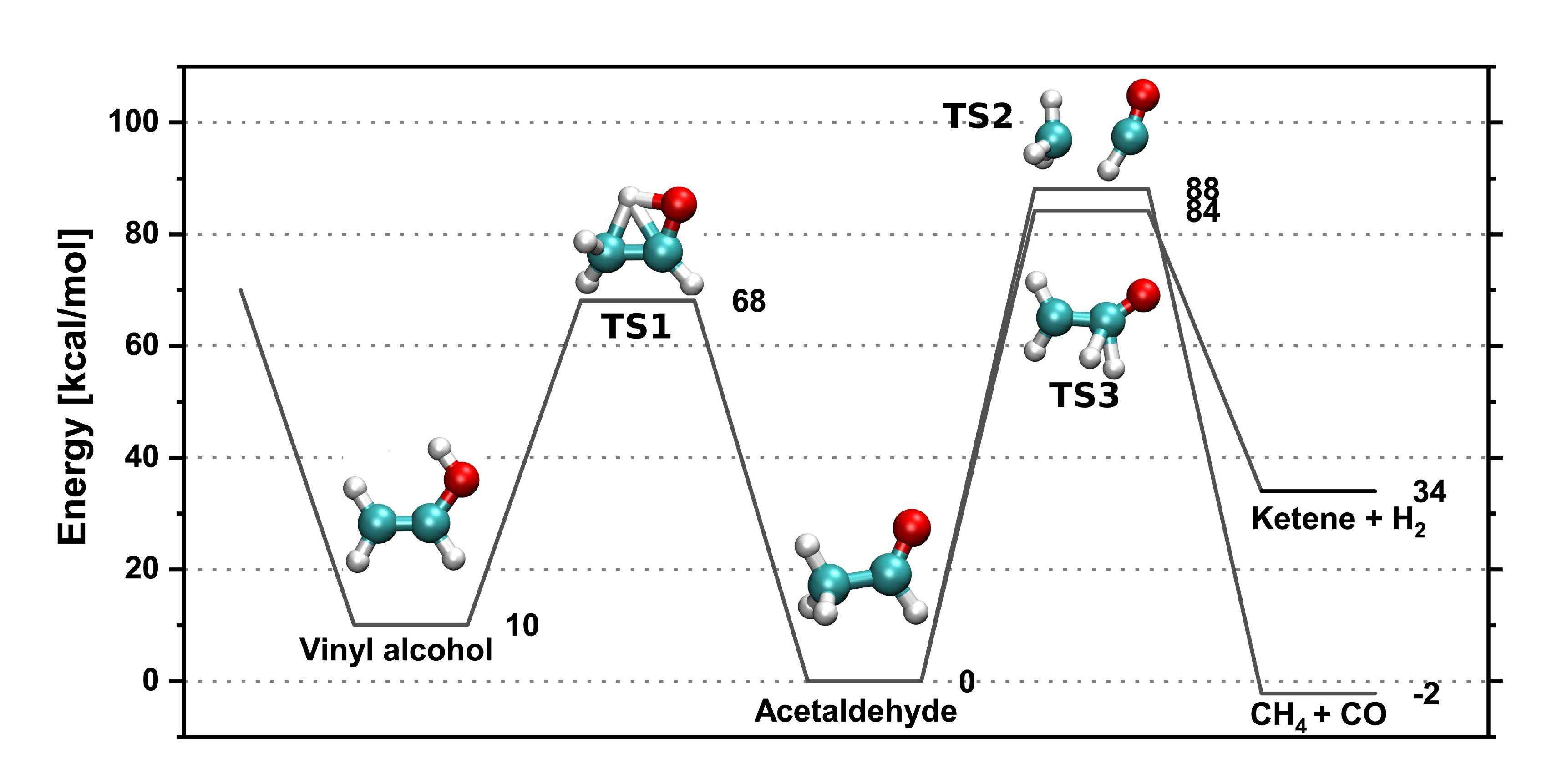}
\caption{Schematic representation of the PES predicted by the NN based
  on MP2/aug-cc-pVTZ reference calculations for the three reactions:
  the tautomerization (to the left) and the two dissociation reactions
  (to the right). All energies are referenced to that of AA. The
  optimized structures and the transition states are illustrated, and
  the corresponding energies are displayed.}
\label{fig:energydiagram}
\end{figure}

\noindent
The energies of the tautomerization and dissociation products from
using the NN (referenced to the optimized structure of AA) are
summarized in Figure \ref{fig:energydiagram}. Tautomerization to VA,
which is 10.1 kcal/mol higher in energy than AA, involves an
activation energy of $\sim 68$~kcal/mol. The two dissociation
reactions have a similar barrier height (to within $\sim
4$~kcal/mol). Table~\ref{table:energies} confirms that the NN performs
well and captures the reference MP2 calculations to within 0.2
kcal/mol. For comparison, energies are also reported from CCSD(T)
calculations which are within 1.0 kcal/mol of the MP2 values except
for the CH$_4$+CO dissociation channel which is less relevant for the
present work.\\

\begin{table}[!h]
\centering
\begin{tabular}{c||c|c|c|c|c|c|c}
 [kcal/mol]& VA & TS1 & AA & TS2 & CH$_4$ + CO & TS3 &ketene + H$_2$ 
\\\hline\hline
$E_{\rm NN}$ & 10.1 & 68.1 & 0 & 88.2 &  --2.2 & 84.2 & 33.8\\
$E_{\rm MP2}$ & 10.1 & 68.1 & 0 & 88.2 &  --2.2 & 84.2& 34.0 \\ \hline
$E_{\rm CCSD(T)}$ & 10.6 & 67.7 & 0 & 83.3 & --6.0 & - & - \\
\end{tabular}
\caption{Comparison of the MP2/aug-cc-pVTZ reference and NN-predicted
  energies for the processes considered. Differences are $\mathbf{\leq
    0.2}$ kcal/mol. For completeness the energies are compared to
  CCSD(T) energies from Reference~\citenum{shaw2018photo}. The zero of
  energy is the optimized structure of AA.}
\label{table:energies}
\end{table}

\subsection{Minimum Dynamical Path}
\label{sec:minimum_dynamical_path}
\begin{sidewaysfigure}
\centering
\includegraphics[width=0.8\textwidth]{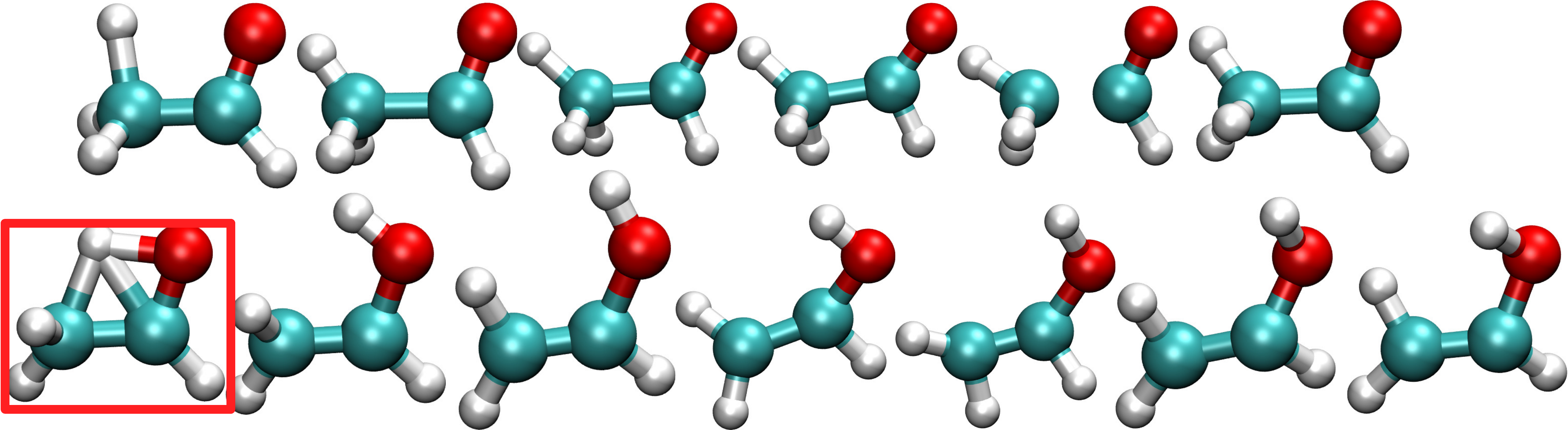}
\caption{MDP for the tautomerization of AA to VA.  The TS is marked
  with a red border.}
\label{fig:mdptauto}
\centering
\includegraphics[width=1\textwidth]{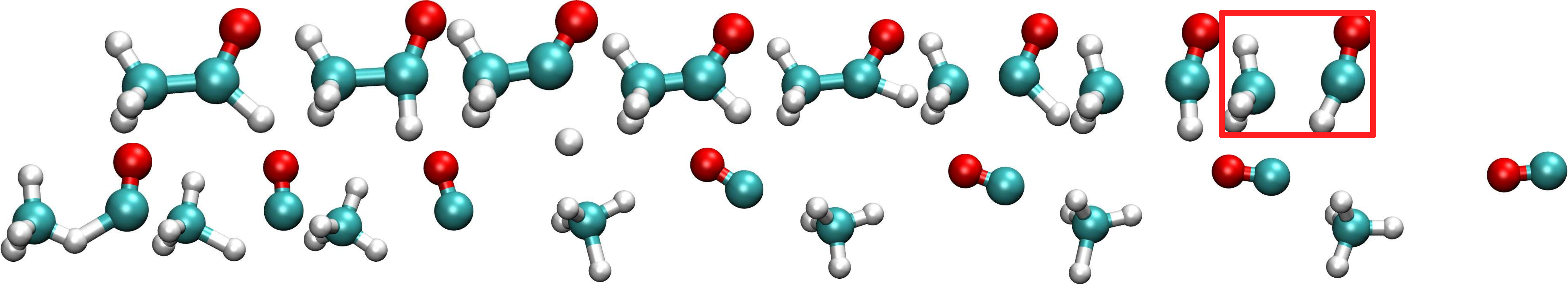}
\caption{MDP for the dissociation of AA to methane and carbon
  monoxide. The TS is marked with a red border.}
\label{fig:mdpdiss}
\centering
\includegraphics[width=1\textwidth]{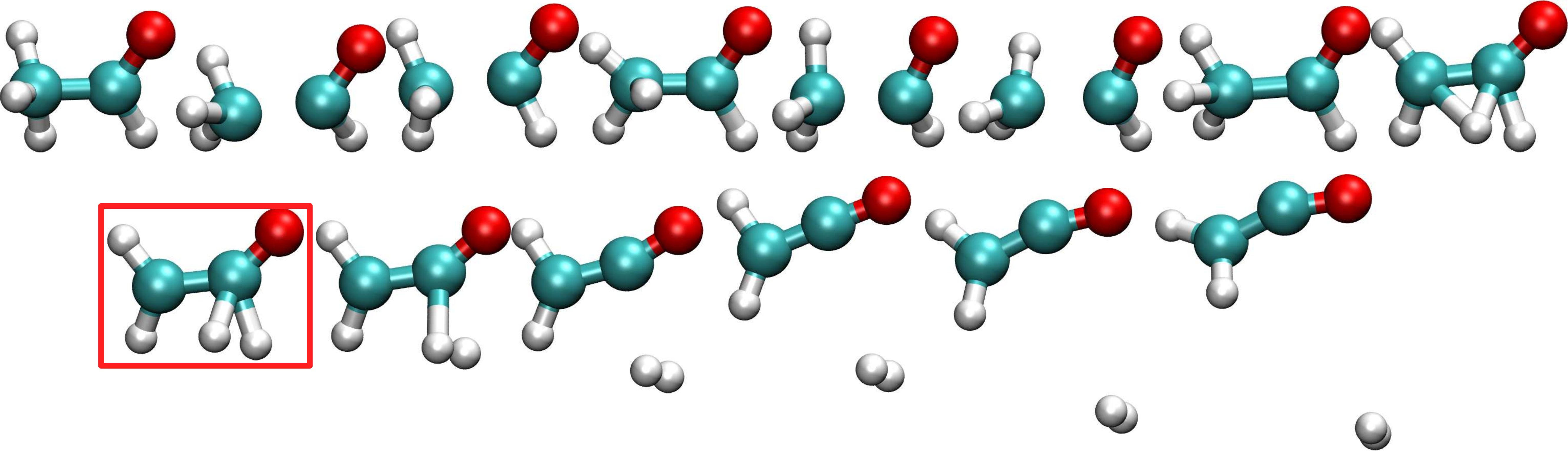}
\caption{MDP for the dissociation of AA to H$_2$ and ketene. The TS is
  marked with a red border.}
\label{fig:mdpketene}
\end{sidewaysfigure}

Structures along the MDPs for the three processes considered here are
shown in Figures~\ref{fig:mdptauto} to \ref{fig:mdpketene}. The TS is
marked with a red border and connects reactant (AA) and product (VA,
CH$_4$+CO, or OC$_2$H$_2$+H$_2$) states from left to right.  For the
tautomerization reaction (Figure \ref{fig:mdptauto}) the transferring
H--atom is involved in a highly excited C--H vibration (see
Figure~S4, green line for $t < 0$) with large
amplitude motion along the CCH angle (Figure~S4,
red line for $t < 0$) which is also coupled to the OH distance
fluctuation before reaching the TS (Figure~S4, blue
line for $t < 0$). Shortly before the TS the H-atoms not involved in
hydrogen transfer (HT) rotate slightly (see Figure~\ref{fig:mdptauto})
which reduces both, the CCH and the CCO angles. After HT the O--H bond
is highly vibrationally excited (Figure~S4, blue
line for $t > 0$) and coupled with an out-of-plane rotation of the
O--H bond (last 3 snapshots of Figure~\ref{fig:mdptauto}). Moreover,
upon hydrogen transfer the CCH and CCO bending vibrations are
out-of-phase by $\pi$.\\

\noindent
For the CH$_4$+ CO dissociation the MDP involves a large-amplitude CH
vibration (Figure~S5, blue line $t < 0$) and a
pronounced oscillation of the OCH bend (between $75^{\circ}$ and
$170^{\circ}$, Figure~S5, black line) while the
methyl rotation is locked, see Figure~\ref{fig:mdpdiss}. For the HT
from the COH moiety to the CH$_3$ group to occur, the CC bond needs to
stretch and be accompanied by OCH bending (peak in OCH angle, see
Figure~S5). Upon dissociation ($t > 0$) the
transferred H-atom exhibits a highly excited CH
(Figure~S5, green line) vibration and the
dissociation products are rotationally excited due to the anisotropy
of the potential energy surface.\\

\noindent
For the dissociation to H$_2$+H$_2$CCO the MDP (see Figure
\ref{fig:mdpketene}) shows that the CC and CH bonds involving the
dissociating H--atoms, which eventually combine to form H$_2$, are
highly vibrationally excited. Moreover, an umbrella motion of the
hydrogen (strong oscillation of the CCH angle between $75^{\circ}$ and
$150^{\circ}$ (Figure~S6, black line) facilitates
the approach of the two H-atoms. After crossing the TS the H$_2$ gains rotational as
well as translational energy whereas the kinetic energy of the ketene is
visible in a bending of the CCO angle (Figure~S6, red line).\\

\noindent
For completeness, structures along the minimum energy path (MEP) are
also reported in Figures~S7 to
S9. Compared to the MDP, the MEP lacks important
dynamic information. For example, the oscillation of the CCO angle in
the tautomerization does not appear in the MEP but plays an essential
role in the MDP.\\

\subsection{Molecular Dynamics Simulations}
\label{sec:results_molecular_dynamics_simulations}
First, the trajectories run at conditions representative for
electronic excitation by solar photons and recent
experiments\cite{shaw2018photo} (\textit{EX} trajectories) are
analyzed. Here, the excitation energy is 93.6 kcal/mol which compares
with energies of 86.6 kcal/mol to 95.3 kcal/mol for actinic
photons. All \textit{EX} trajectories starting from the optimized AA
structure remained in the AA well and none of the 2000 trajectories
exhibited isomerization to VA. In other words, starting from the
structure after photoexcitation, IVR is very efficient and localizes
the system in the AA structure. Although the energized AA contains
sufficient energy to isomerize to VA ($\sim 94$ kcal/mol compared with
a barrier height of 68 kcal/mol for isomerization to VA), no such
process is found after a total simulation time of 500 ns. This
suggests that on the $\sim 100$~ns time scale isomerization
AA$\rightarrow$VA is unlikely for excitation energies compatible with
actinic photons.\\

\begin{figure}[!hb]
\centering
\includegraphics[width=0.9\textwidth]{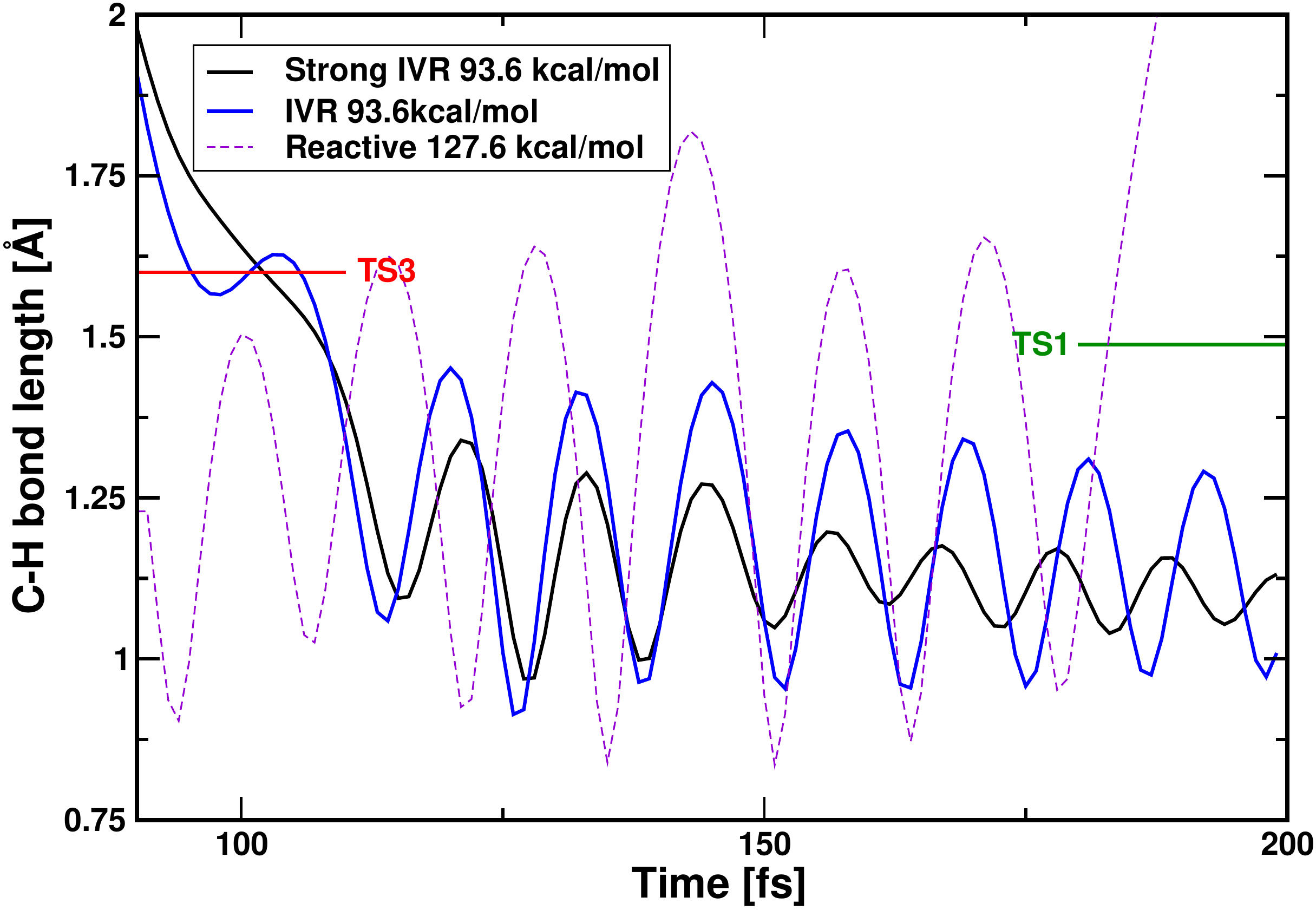}
\caption{The time series for the CH stretch for three different
  trajectories. Two trajectories are started from TS3 with an energy
  of 93.6~kcal/mol with respect to the ground state AA (blue and
  black). For the black trace the motion along the CH mode is strongly
  damped. The red line indicates the CH bond length of TS3 and the
  green line corresponds to TS1 connecting AA and VA (see
  Figure~\ref{fig:energydiagram}) which needs to be reached for
  AA$\rightarrow$VA isomerization. The purple line corresponds to the
  CH stretch of a reactive ZPE trajectory undergoing AA$\rightarrow$VA
  reaction.}
\label{fig:nn_ch_time_series}
\end{figure}

\noindent
To further investigate the role of IVR, five independent trajectories
were started from TS3 with an excess energy of 93.6 kcal/mol with
respect to the ground state of AA. The analysis of these trajectories
relaxing from TS3 towards AA reveals that the energy is rapidly (few
tens of fs) distributed among the different modes (see
Figure~\ref{fig:nn_ch_time_series}). In order to validate these
findings, separate {\it ab initio} MD simulations were run using the
semiempirical tight binding GFN2-xTB\cite{bannwarth2019gfn2}
method. These simulations support that IVR of the relevant CH stretch
modes occur on the sub-ps time scale, see
Figure~S10. Hence, irrespective of system
preparation and energy function used, isomerization from AA to VA is
not found in the present simulations on the aggregate, multiple 100 ns
simulations when excitation energies consistent with actinic photons
are used.\\

\noindent
Next, 10'000 \textit{ZPE} trajectories were run for an excitation
energy of 127.6~kcal/mol, as had been done in earlier computational
work for AA decomposition into CH$_4$+CO.\cite{bowman:2007} The error
between two independently trained NNs (see
Figure~S2) for a typical trajectory is
0.047~kcal/mol with a standard deviation of 0.074~kcal/mol. This
indicates that the NN used in the MD simulation is robust. It is
worthwhile to note that at an earlier stage of the NN with a total
(i.e.\ training, validation, and test) of 411'204 structures none of
the trajectories run at an excitation energy of 127.6 kcal/mol
fulfilled such a tight statistical criterion. Hence, an additional
21'195 structures was specifically added in the high-energy part of
the reference data set. Without this, the present quality of the
simulations could not have been achieved.\\

\noindent
Secondly, the trajectories were categorized based on the type of
reactions that occur. Out of the 10'000 runs, 9344 trajectories are
unreactive -- i.e.\ remain in the AA structure, as was found for the
lower excitation energy considered above -- 26 trajectories dissociate
to CH$_4$+CO, 22 react to ketene+H$_2$, and a total of 608 isomerize
to VA. Hence, by using a considerably larger excitation energy than
that available from solar photons the process of interest --
isomerization from AA to VA -- is observed on the $\sim 100$~ps time
scale for 6 \% of the trajectories. The reaction times, defined as the
time difference between the start of the trajectory and reaching an
OH-separation of $< 1$~\AA\/ for the first time, are summarized in
Figure~S11. The reaction time distribution is flat and
includes everything from prompt isomerization (few ps) to exhausting
the total simulation time (50 ps).\\

\begin{figure}[!hb]
\centering
\includegraphics[width=0.9\textwidth]{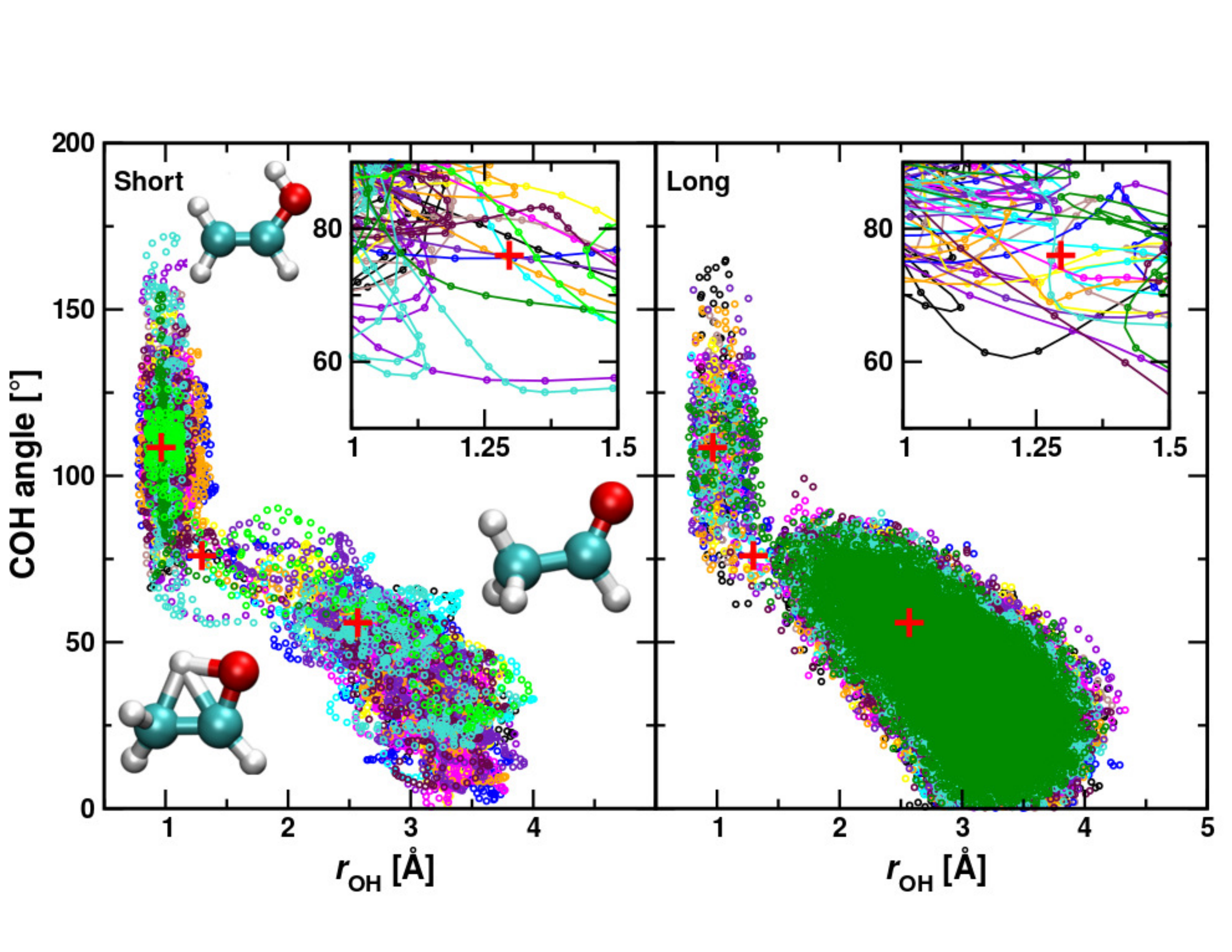}
\caption{Projection of the reaction path from AA to VA in terms of the
  O--H bond length $r_{\rm OH}$ and COH bond angle for the
  trajectories with the 13 shortest and longest reaction times. Trajectories 
  with short reaction times are shown up to 200~fs
  after the tautomerization. The optimized structures for AA, for VA
  and the TS are marked with a cross, and illustrated with the
  corresponding structures in the left panel.}
\label{fig:zpe_scatter}
\end{figure}

\noindent
The isomerization dynamics from AA to VA is further analyzed by
considering the reaction path as a function of the OH distance and the
COH angle, see Figure~\ref{fig:zpe_scatter}, which are two of the main
coordinates involved in the process. Here, the AA structure is in the
lower right-hand corner whereas that of VA is in the upper left hand
corner. For short reaction times (left panel in
Figure~\ref{fig:zpe_scatter}) the AA well is scarcely sampled whereas
for trajectories with long reaction times (right panel), exhibiting
partial IVR, this is not the case. In this projection, TS1 is located
in (or very close to) the region covered by the VA well. Therefore,
the two coordinates chosen do not best separate reactants from
products, in particular with respect to the $r_{\rm OH}$
coordinate. The distribution of the reactive trajectories in the
region around the TS has a comparable width along the COH coordinate
(see insets) although short-lived trajectories (short reaction times)
sample angles down to $55^\circ$ compared to $60^\circ$ for
longer-lived trajectories. The most apparent difference is that the
short-lived trajectories cross the TS in a direct manner whereas the
long-lived trajectories also show recrossings (e.g.\ green) before
reaching the VA well. This is consistent with a ballistic mechanism
for the short-lived and a structurally more heterogeneous dynamics for
long-lived trajectories. As a consequence, TST-based approaches will
have difficulties to correctly capture the reaction rate.\\

\noindent
Next, the trajectories are evaluated after tautomerization has
occurred. Out of 608 trajectories that tautomerize to VA, 31
re-isomerized to AA which are unlikely to proceed further to FA
(i.e.\ cross again from AA to VA) but a thorough assessment of this is
outside the scope of this work and would require considerably more and
longer sampling. Furthermore, it is found that $\sim 50$~\% of the
re-isomerizing trajectories (VA$\rightarrow$AA) access regions in
configurational space poorly covered by the reference \textit{ab
  initio} calculations which is reflected in occasional breaking of
the C-O bond.\\

\noindent
As soon as the system has isomerized to VA the quality of the dynamics
deteriorates. This is illustrated by considering a trajectory which
isomerizes (AA$\rightarrow$VA) and then samples the VA well
extensively. Visual inspection of the trajectory reveals that the
OH-group temporarily dissociates but eventually rebinds to the carbon
atom again.  This trajectory is evaluated with two independently
trained NNs and the energy predictions are compared in
Figure~S2. The deviation between the two NNs
shows one large spike when the OH moiety dissociates. Except for this,
only modest differences between the two NNs are found. However, for a
comprehensive description of the dynamics in the VA-well, additional
structures would be required and the NN would have to be retrained. As
this part of the dynamics is not of direct interest in the present
work this was, however, not done.\\

\section{Discussion and Conclusion}
Using an NN trained on a large number of reference data provides an
accurate, fully-dimensional and reactive PES describing AA, VA, and
the dissociation to CH$_4$+H$_2$ and H$_2$+H$_2$C$_2$O, see Figure
\ref{fig:corracetaldehyde}. The PES is suitable to run reactive MD
simulations and to analyze several reaction pathways. However, when
running simulations with sufficiently high excitation energies to
induce isomerization between AA and VA, validation of the sampled
structures and energies is mandatory. In generating the NN-based PES
it was found that when training on a slightly smaller data set
(containing a total of 411'204 structures) the NN was not suitable for
robust simulations of the AA$\rightarrow$VA isomerization at 127.6
kcal/mol and additional reference calculations had to be included to
cover the high-lying regions of the PES. Histograms showing different
bond lengths for the full data set and the difference between this and
the smaller data set are provided in Figures~S12 to
19.\\

\noindent
The accuracy of the fitted PES can also be discussed by considering
MAEs and RMSEs in specific energy intervals relevant to the processes
of interest in the present work, see
Table~\ref{tab:test_set_energyranges}. For energies below the
isomerization barrier to VA (68 kcal/mol) the MAE and RMSE are
0.0081~kcal/mol and 0.0131~kcal/mol, respectively. This remains
similar for energies up to the energy available in the solar spectrum
($\sim 93.6$ kcal/mol) where the MAE is 0.0071~kcal/mol and the RMSE
is 0.0145~kcal/mol. For energies corresponding to the highest
excitation energy used in the simulations (127.6 kcal/mol) the MAE and
the RMSE increase slightly to 0.0132~kcal/mol and 0.0307~kcal/mol,
respectively. Finally, for energies larger than 127.6~kcal/mol, a MAE
of 0.1226~kcal/mol and an RMSE of 2.0728~kcal/mol are found. A system
for which such data has been published is N$_2$+N$_2$ for which
typical RMSEs of a parametrized fit are 1.8~kcal/mol (for energies up
to 100~kcal/mol) and 4.1~kcal/mol (for energies up to 228~kcal/mol),
respectively.\cite{paukku2013global}\\

\begin{table}[h]
\begin{tabular}{l|ccc}
$E$ [kcal/mol] & \#   & MAE($E$) & RMSE($E$) \\\hline
$E< 30$         & 5550 & 0.0045 & 0.0067  \\
30 $< E <$ 68  & 5719 & 0.0081 & 0.0131  \\
68 $< E<$ 93.6  & 4297 & 0.0071 & 0.0145  \\
93.6 $<E <$ 127.6 & 2857 & 0.0132 & 0.0307  \\
$E>$ 127.6        & 8976 & 0.1226 & 2.0728 \\\hline
total & 27399 & 0.0453 & 1.1865
\end{tabular}
\caption{Errors (in kcal/mol) of the fitted PES with respect to the
  \textit{ab initio} MP2/aug-cc-pVTZ calculations for different energy
  ranges. The structures of the test set are evaluated and the energy
  of the optimized AA structure serves as the zero of energy. }
\label{tab:test_set_energyranges}
\end{table}

\noindent
Starting trajectories from the optimized AA structure at 93.6 kcal/mol
above the minimum energy structure of AA does not lead to VA. For
this, an aggregate of 0.5~\textmu s of reactive MD simulations was run
and analyzed. It is, therefore, concluded that nonequilibrium
preparation of AA through absorption of an actinic photon $\lambda =
300$ to 330 nm (86.6 to 95.3 kcal/mol) isomerization to VA on the
sub-$\mu$s time scale is unlikely to occur.\\

\noindent
The present PES can also be used to study decomposition into CH$_4$+CO
and H$_2$+H$_2$CCO which is, however, outside the scope of the present
work. Previous efforts included simulations for the AA $\rightarrow$
CH$_3$+HCO and the AA $\rightarrow$ CH$_4$+CO dissociation from {\it
  ab initio} MD
simulations\cite{kurosaki2003photodissociation,kurosaki2002photodissociation,
  kurosaki2006photodissociation}, for the AA $\rightarrow$
H$_2$CO+H$_2$ dissociation on a reaction path
potential\cite{harrison2019dynamics} and QCT simulations on a global
ab initio-based PES fitted using permutationally invariant polynomials
(PIPs).\cite{bowman:2012} Additionally, the roaming dynamics in the
dissociation of acetaldehyde was examined using a reduced
dimensionality trajectory approach \cite{harding2010roaming} and using
a full-dimensional PES fit with PIPs.\cite{bowman:2007}\\

\noindent
Because the \textit{EX} trajectories yielded no reactive events and it
was found that IVR is an impeding factor to reactivity and
isomerization, RRKM (statistical rate) calculations were also carried
out. Rate constants $k(E)$ are determined with the MultiWell 2016
suite of programs \cite{barker2001multiple}. Key parameters (harmonic
frequencies, energies of the critical points, see
Tables~S1 and S2) are
obtained from the NN (trained at MP2/aug-cc-pVTZ) and the rates were
computed for two different total energies, 93.6 and 127.6 kcal/mol,
corresponding to the two types of MD simulations carried out. The
isomerization reactions AA$\leftrightarrow$VA occur on the ns time
scale, see Table \ref{table:rate_constants} whereas the dissociation
reactions (to CH$_4$+CO and H$_2$+ketene) with energies of 93.6
kcal/mol are predicted to occur on the $\mu $s time scale. With an
energy of 127.6~kcal/mol both decomposition reaction rates are on the
ns time scale.\\

\noindent
Such rates are consistent with earlier work. At the B3LYP level of
theory the RRKM rate for
photodissociation\cite{gherman2001photodissociation} of AA to
CH$_4$+CO at 121.76~kcal/mol (i.e.\ 5.28~eV) $k(E)=4.47\times10^9$
compared with $k(E)=1.715\times10^9$ from the present calculations
using the NN-trained PES. Similar work considering statistical rates
has been done on AA. It includes the dissociation of AA following
excitation into S$_1$ \cite{tachikawa1994photodissociation} or the
FA-assisted tautomerization of AA\cite{peeters:2015}.\\

\begin{table}[!h]
\centering
\begin{tabular}{c||c|c}
\textbf{$k(E)$ [s$^{-1}$]} & 93.6 kcal/mol & 127.6 kcal/mol \\ 
\hline\hline
VA $\rightarrow$ AA& $7.209\times 10^8$ & $2.073\times 10^{10}$\\
AA $\rightarrow$ VA&  $1.569\times 10^8$ & $5.865\times 10^9$\\
AA $\rightarrow$ H$_2$+H$_2$CCO & $2.612\times 10^6$ & $1.098\times10^9$\\ 
AA $\rightarrow$ CH$_4$+CO& $1.699\times 10^6$ & $3.756\times10^9$
\end{tabular}
\caption{Rates [s$^{-1}$] for the possible AA reactions at two
  different total energies used in the MD simulations.}
\label{table:rate_constants}
\end{table}

\noindent
On the other hand, running simulations including zero-point
vibrational energy at a total energy of 127~kcal/mol readily leads to
isomerization on the 100 ps time scale (608 events in 500 ns). The
fact that for the dynamics in the VA well the NN may have deficiencies
does not affect this conclusion because up to TS1, separating AA from
VA, the quality of the NN is very good.\\

\noindent
The flat reaction time distribution for isomerization between AA and
VA differs from reaction time distributions for water elimination from
H$_2$SO$_4$ after vibrational excitation of an overtone OH-stretch
vibration\cite{MM.h2so4:2011,MM:2014.h2so4} where they followed a
Poissonian distributions. Since in the present case the initial
preparation is thermal and not mode specific and IVR is very rapid it
is conceivable that the distribution of reaction times is flat. In
other words, after relaxing into the AA well, the energy (here 127.6
kcal/mol) is either used to isomerize in a ballistic manner or
isomerization to VA occurs after partial relaxation on longer time
scales. For the smaller excitation energy (93.6 kcal/mol) only
relaxation into AA was observed on the time scale of the simulations
(aggregate of 500 ns).\\

\noindent
Whether or not isomerization of AA to VA in the atmosphere is relevant
for FA generation depends, therefore, on the competition between IVR
and collisional de-excitation. Under upper stratospheric-lower
mesospheric conditions the deactivation of excited molecules due to
collisions with other molecules are relevant. For H$_2$SO$_4$ at
typical polar (75$^\circ$S) conditions in the stratopause (at a height
of $50$ km with pressure $p \sim 70$ Pa and temperature $T \sim 280$
K) quenching occurs on the $130$ ns timescale (inverse of the
collision frequency $Z$,
$\tau_\textrm{quench}=(Z[\textrm{M}])^{-1}$).\cite{MM:2014.h2so4}
Given their similar mass and size, H$_2$SO$_4$ and AA are expected to
behave similarly and it is expected that this time scale is also
representative for AA. If after photoexcitation and returning to the
ground state AA has not isomerized (to VA) or decayed (to products)
within $\sim 100$ ns, it is likely that de-excitation occurs through
collisions. Further chemical processing, including isomerization and
decomposition, is not possible then. Hence, formation of FA following
electronic excitation with actinic photons of AA and subsequent ground
state relaxation and isomerization to VA appears unlikely to occur.\\

\section{Acknowledgments}
This work was supported by the Swiss National Science Foundation
through grants 200021-117810, 200020-188724, the NCCR MUST, and the
University of Basel. OTU acknowledges funding from the Swiss National
Science Foundation (Grant No. P2BSP2\_188147).\\

\bibliography{references}
\end{document}


\clearpage
\section{Reference Structure Set}
\begin{figure}[!h]
\centering
\includegraphics[width=1\textwidth]{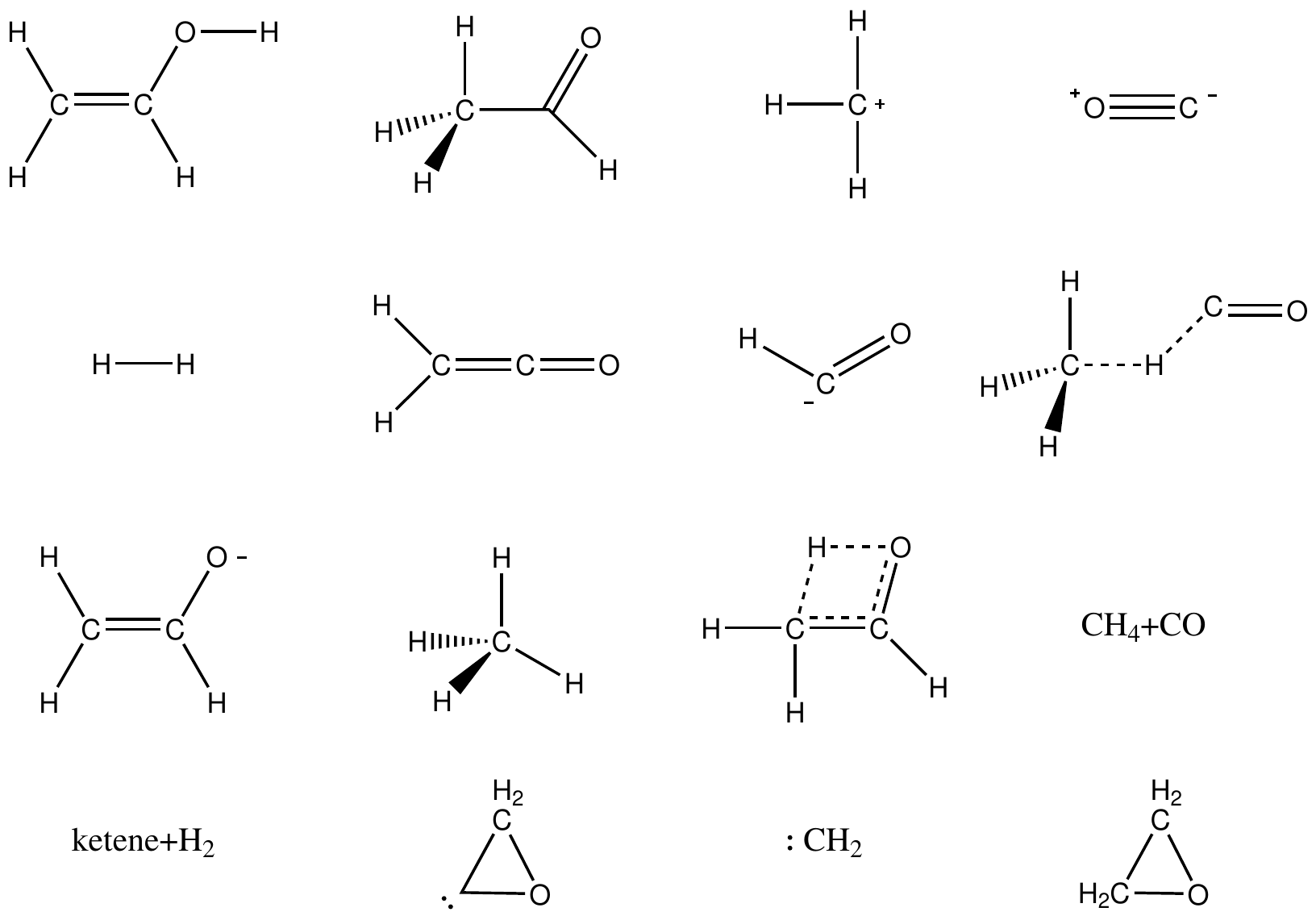}
\caption{Schematic representation of all molecules and compounds
  (``amons'') used in NN training for the acetaldehyde PES.}
\label{sifig:ref_structures}
\end{figure}

\clearpage
\section{NN Energy Comparison}
\begin{figure}[!h]
\centering
\includegraphics[width=1\textwidth]{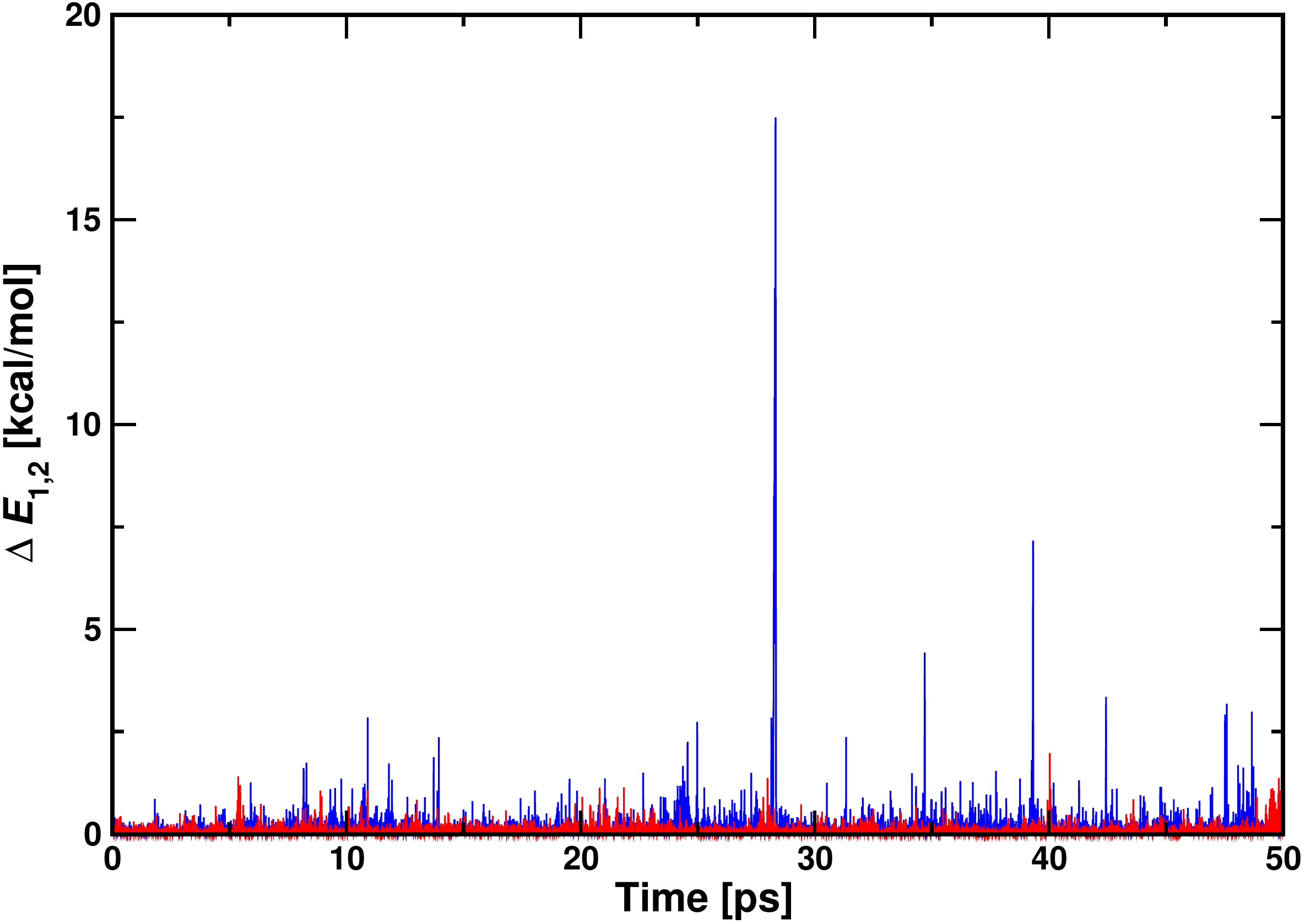}
\caption{Energy deviation between the two independently trained NNs
  for a random unreactive trajectory (red) and a random reactive
  trajectory (blue, containing a tautomerization to VA) run at 127.6
  kcal/mol. The unreactive trajectory has a mean of 0.047~kcal/mol and
  a standard deviation of 0.074~kcal/mol, whereas the reactive
  trajectory has a mean of 0.108~kcal/mol and a standard deviation of
  0.476~kcal/mol. The large spike in the blue curve marks where the OH
  moiety dissociates and illustrates that the NN is not robust.}
\label{sifig:stdev_nn_127}
\end{figure}

\clearpage
\section{Systematic Outliers}
\begin{figure}[ht]
\centering
\includegraphics[width=0.8\textwidth]{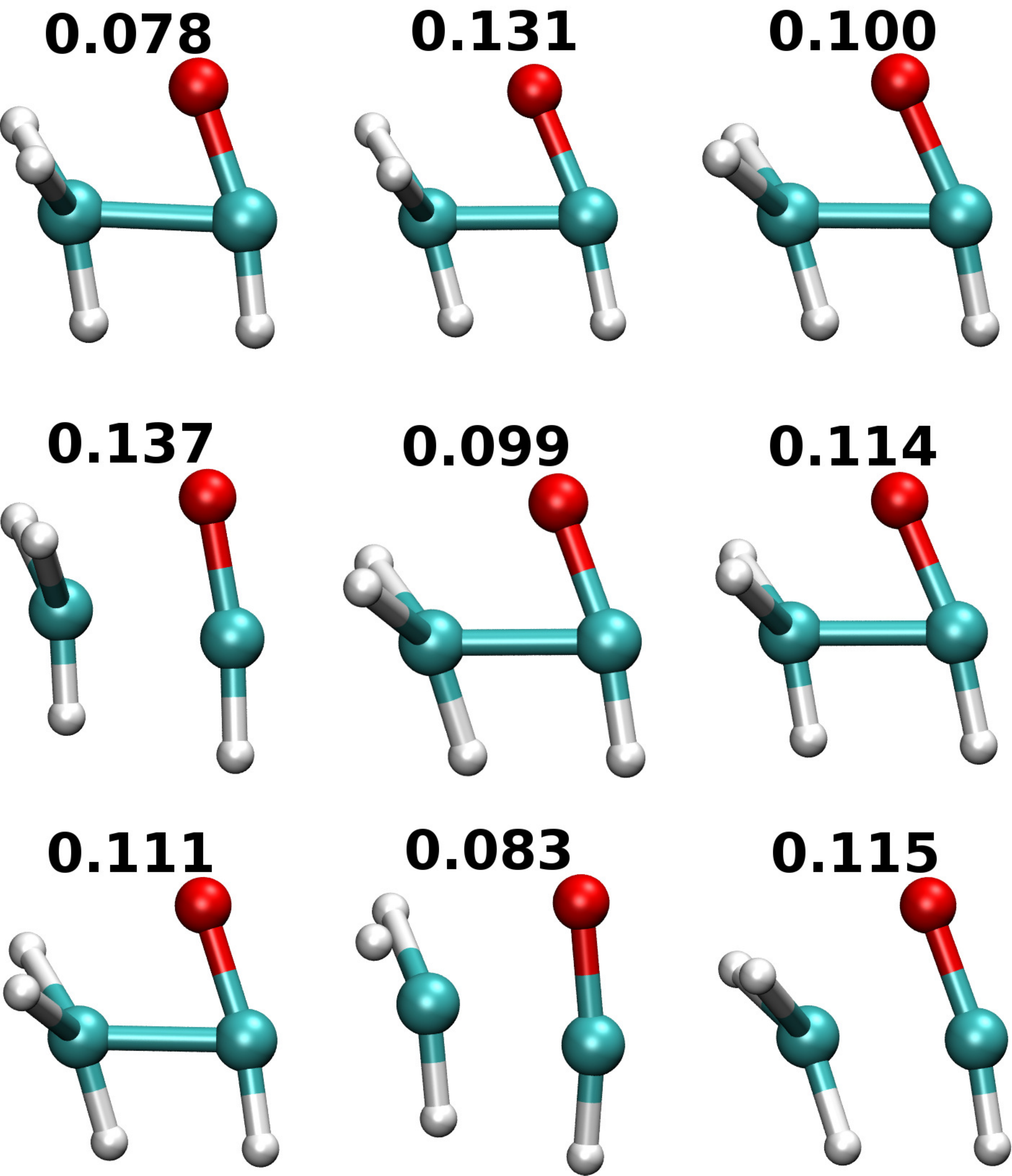}
\caption{Illustration of structures from the test set showing a
  statistical error in the energy prediction of the NN. Moreover, the
  T$_1$ value is shown and bonds are illustrated up to 2.7~\r{A}.}
\label{sifig:stat_error}
\end{figure}

\clearpage
\section{Minimum Dynamic Path}
\begin{figure}[!h]
\centering \includegraphics[width=0.6\textwidth]{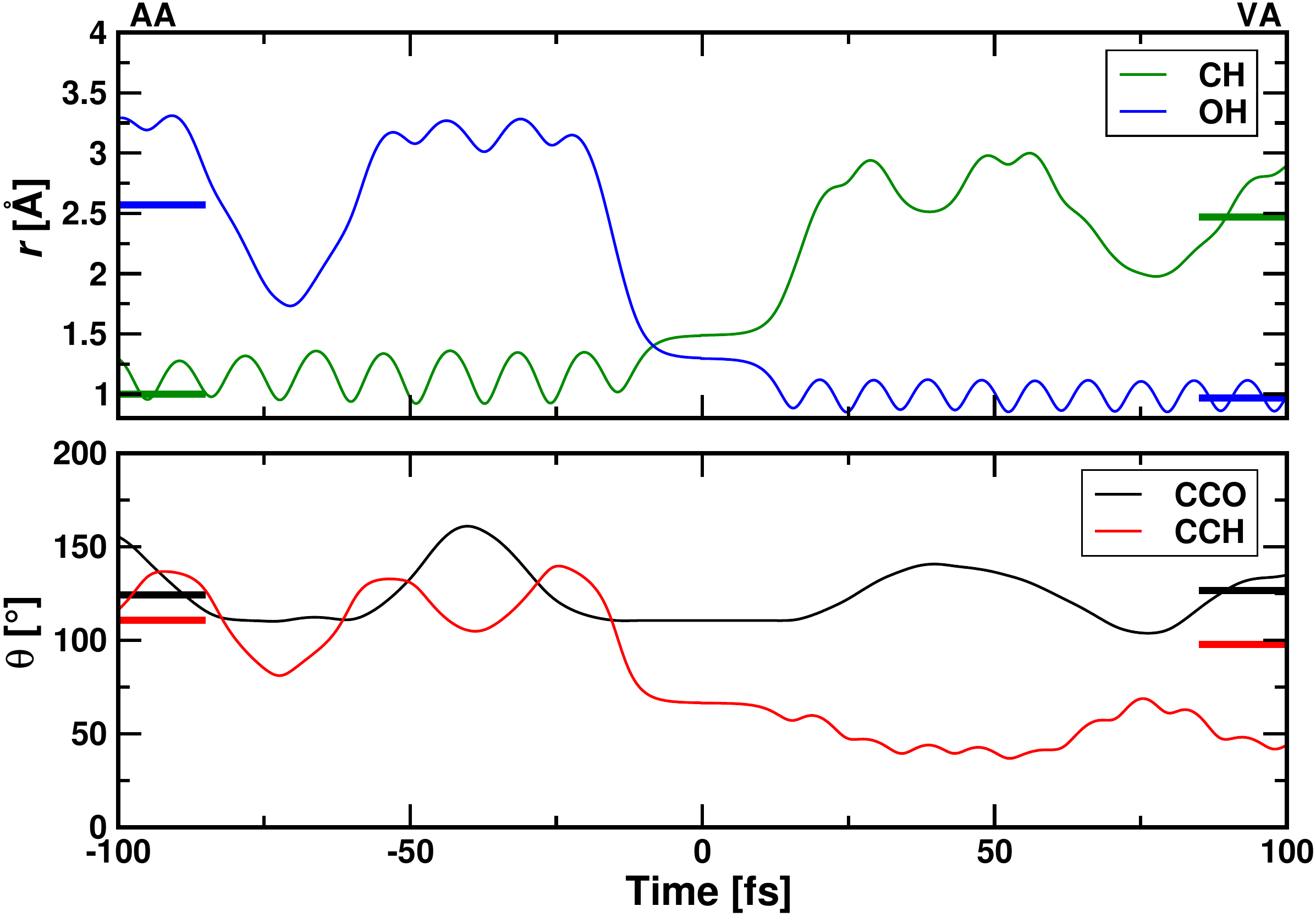}
\caption{Evolution of different geometric properties during the MDP of
  the tautomerization reaction. The bars mark the equilibrium values
  for AA (left) and VA (right).}
\label{sifig:mdp_tauto}
\end{figure}

\begin{figure}[!h]
\centering
\includegraphics[width=0.6\textwidth]{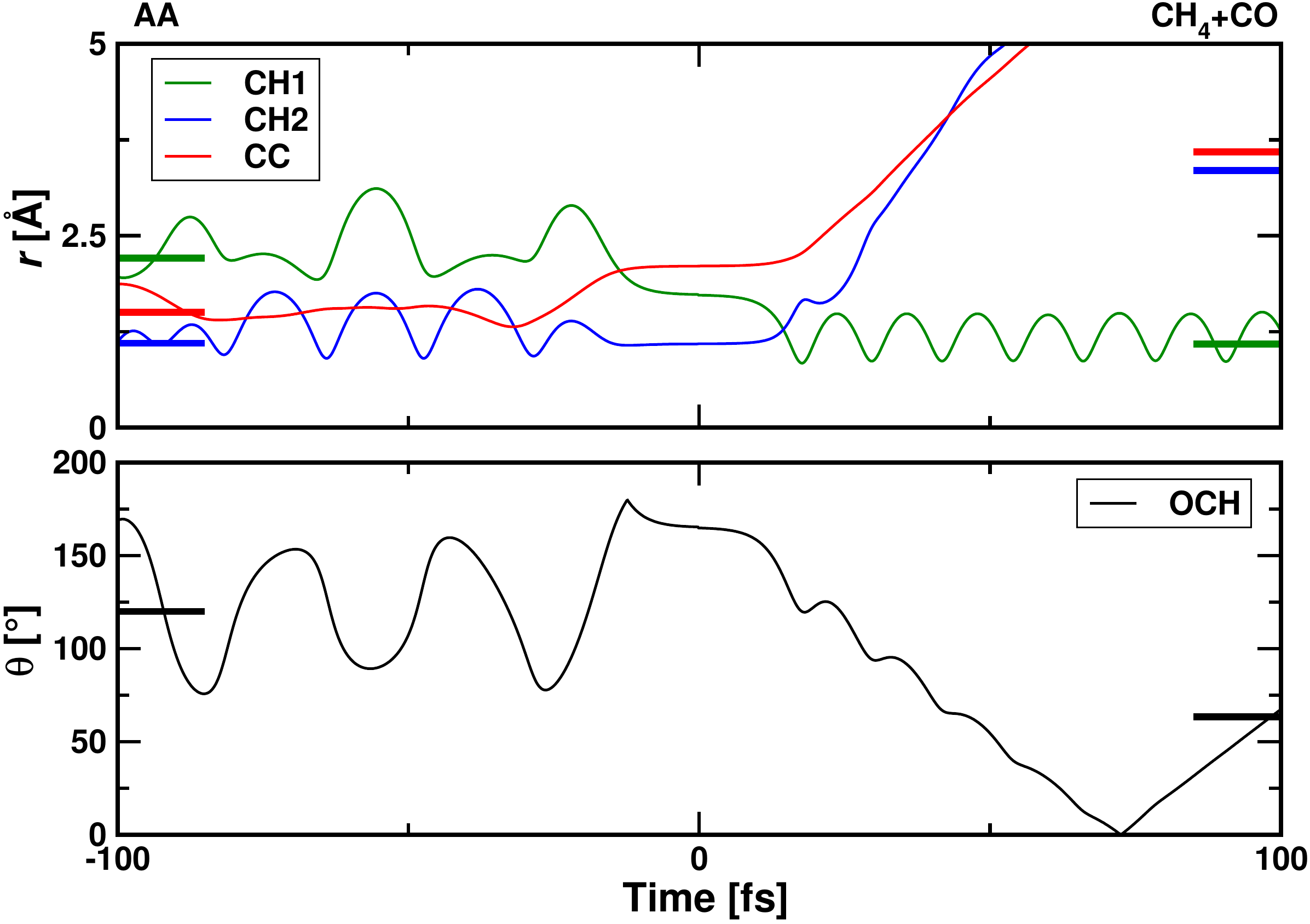}
\caption{Evolution of different geometric properties during the MDP of
  the dissociation reaction forming CO and CH$_4$. The bars mark the
  equilibrium values for AA (left) and the CH$_4$+CO dissociation
  product. }
\label{sifig:mdp_disso}
\end{figure}
\clearpage
\begin{figure}[!h]
\centering
\includegraphics[width=0.6\textwidth]{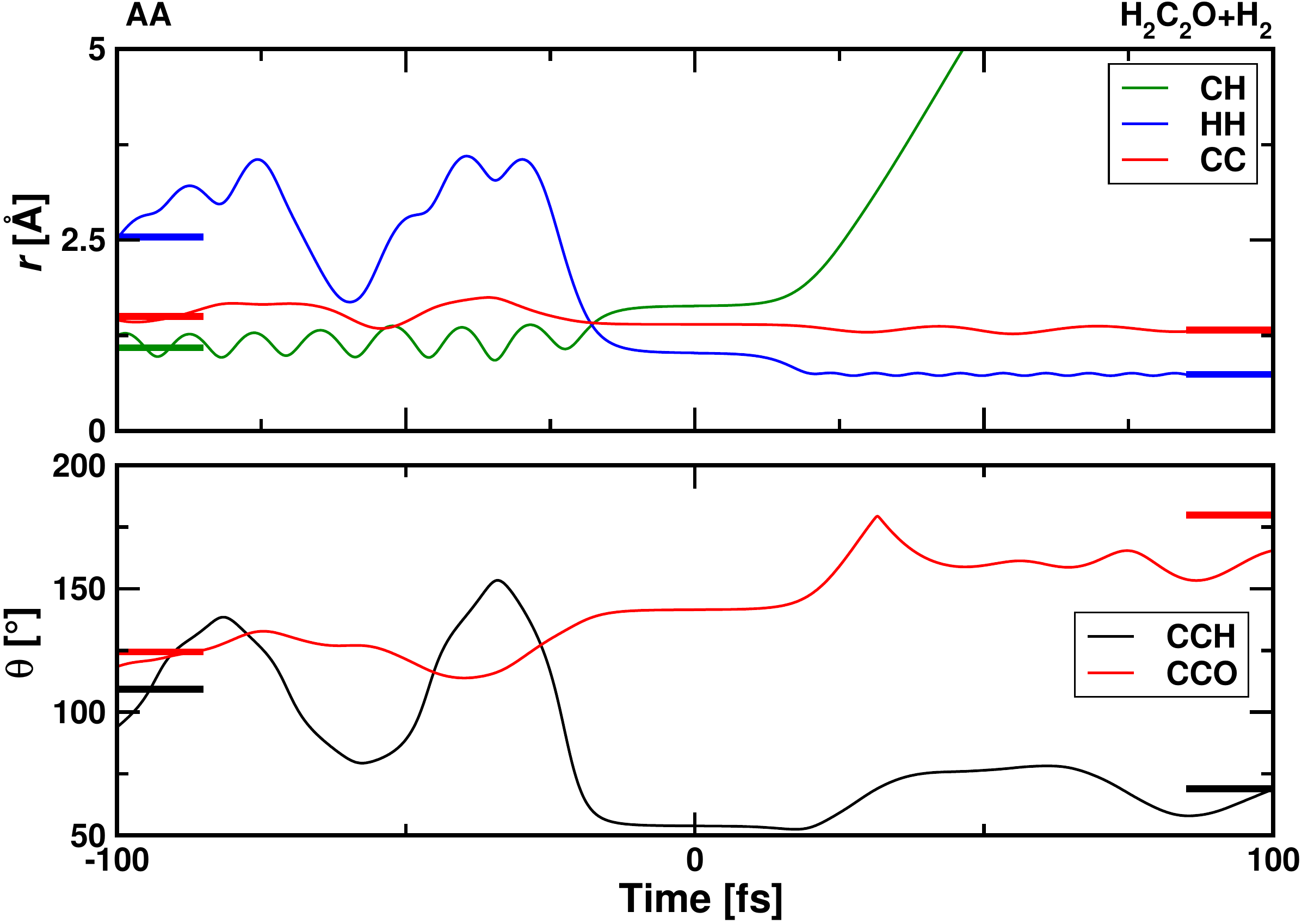}
\caption{Evolution of different geometric properties during the MDP of
  the dissociation reaction forming H$_2$CCO and H$_2$. The bars mark
  the equilibrium values for AA (left) and the H$_2$CCO and H$_2$
  dissociation product.}
\label{sifig:mdp_newts}
\end{figure}
\clearpage
\section{Minimum Energy Path}
\begin{figure}[!h]
\centering
\includegraphics[width=1\textwidth]{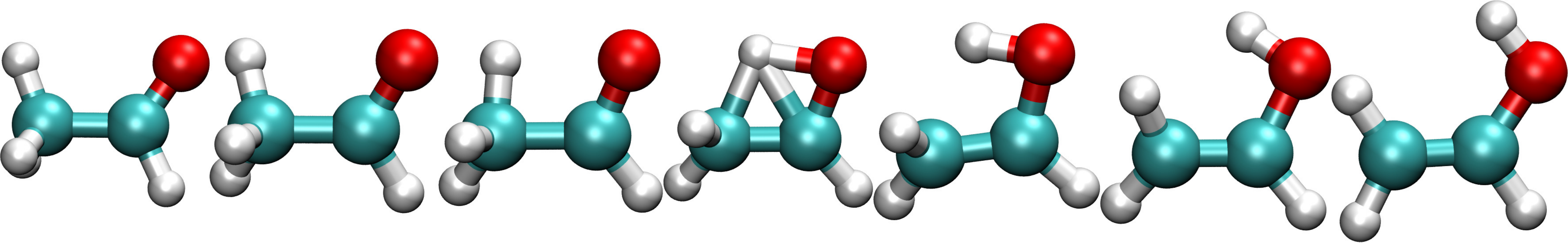}
\caption{MEP for the tautomerization reaction.}
\label{sifig:mep_tauto}
\includegraphics[width=1\textwidth]{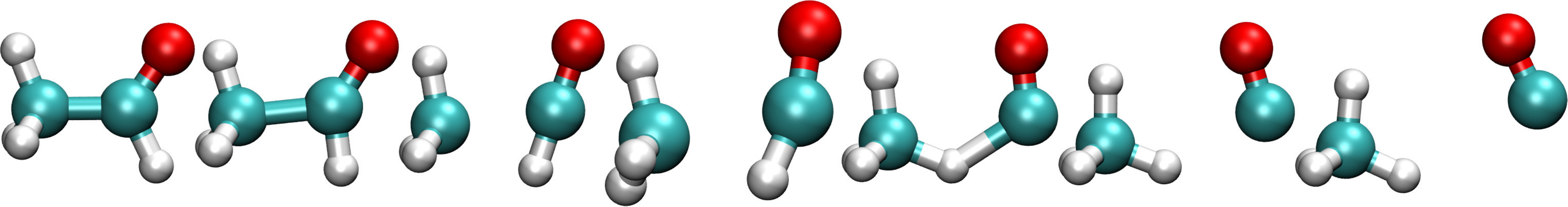}
\caption{MEP for the dissociation to CH$_4$ and CO.}
\label{sifig:mep_diss}
\includegraphics[width=1\textwidth]{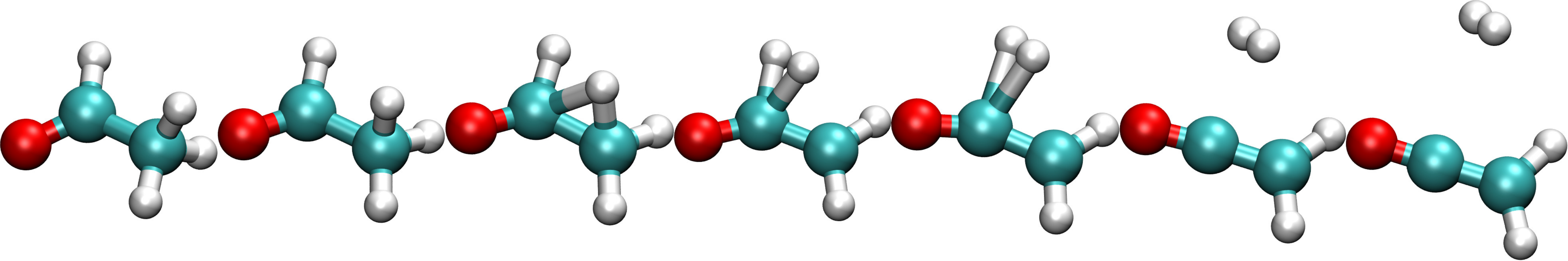}
\caption{MEP for the dissociation to H$_2$ and H$_2$CCO.}
\label{sifig:mep_newts}
\end{figure}
\clearpage
\section{EX Trajectories}
\begin{figure}[!h]
\centering
\includegraphics[width=1\textwidth]{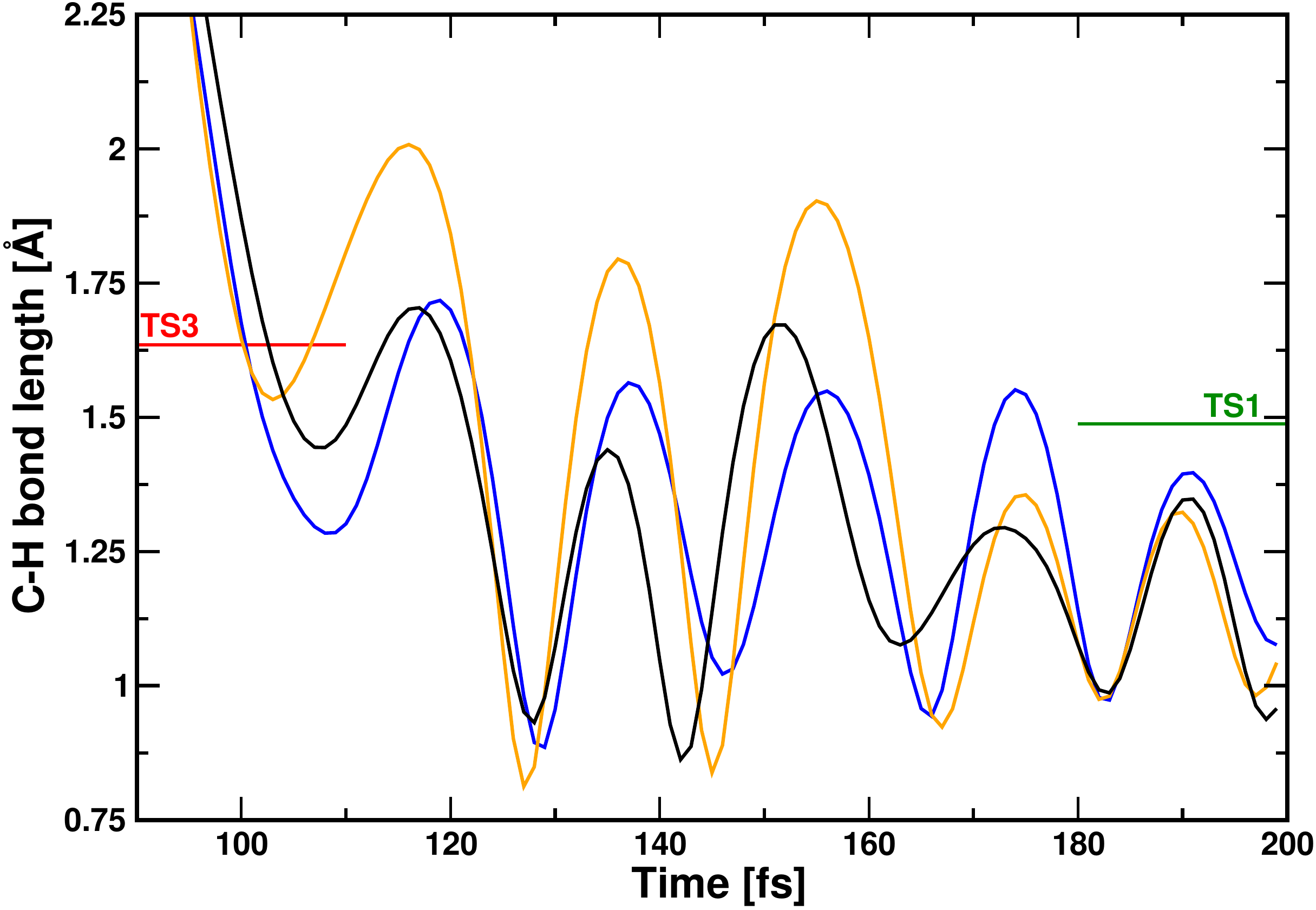}
\caption{The progression in time for the C--H stretch motion is shown
  for three trajectories calculated using the GFN2-xTB method. A rapid
  decrease of the magnitude in C--H stretching is visible after two or
  three oscillations. The red line indicates the C--H bond length of
  TS3 (see Fig.~3).}
\label{sifig:xtb_ch_time_series}
\end{figure}
\clearpage
\section{ZPE Trajectories}
\begin{figure}[!ht]
\centering
\includegraphics[width=0.8\textwidth]{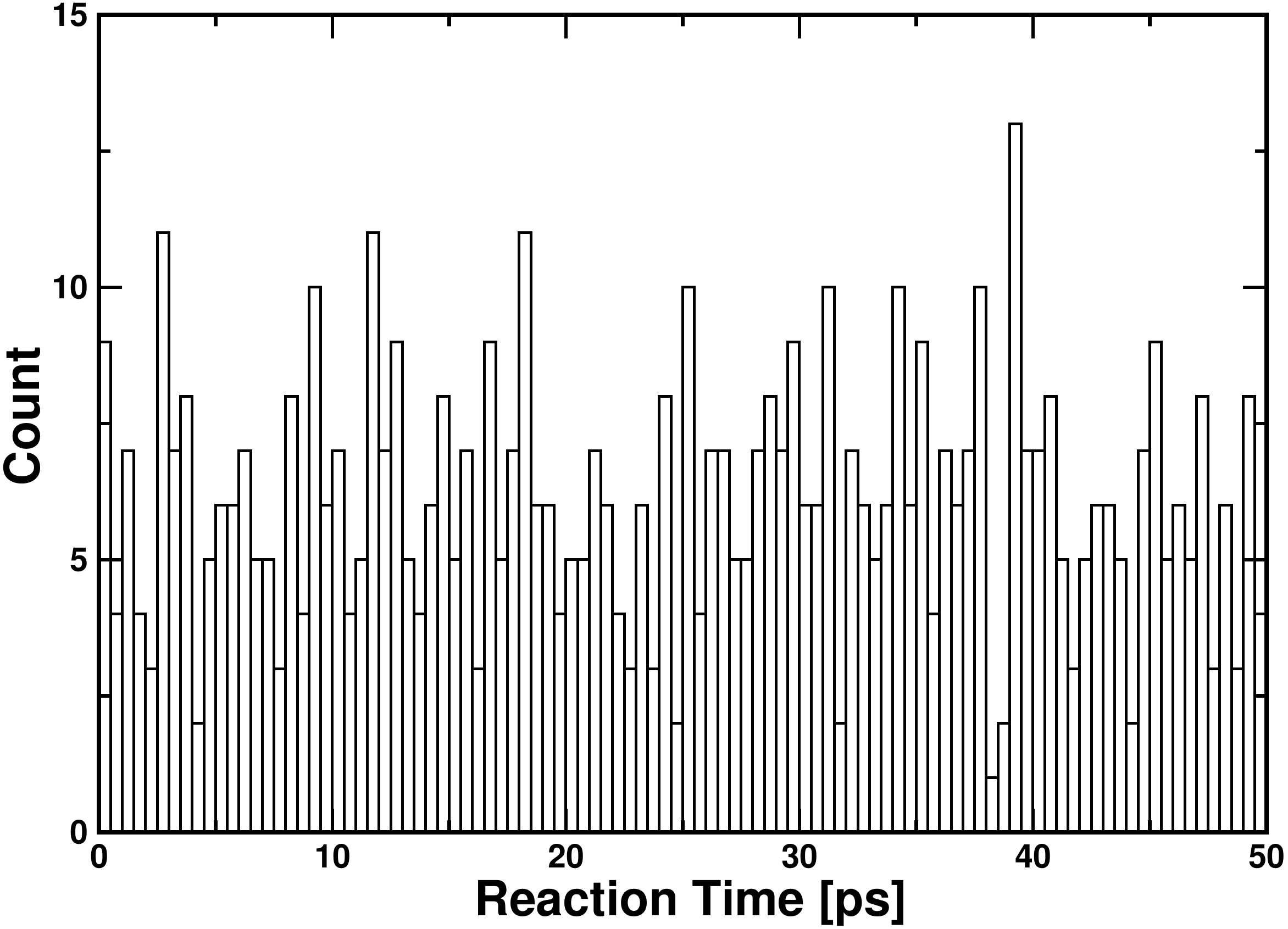}
\caption{Histogram representing the simulation time before the
  tautomerization from AA to VA for a total of 608 tautomerizations.}
\label{sifig:rt_VA}
\end{figure}
\clearpage
\section{Bond Length Distributions}
\begin{figure}[ht]
\centering
\includegraphics[width=0.7\textwidth]{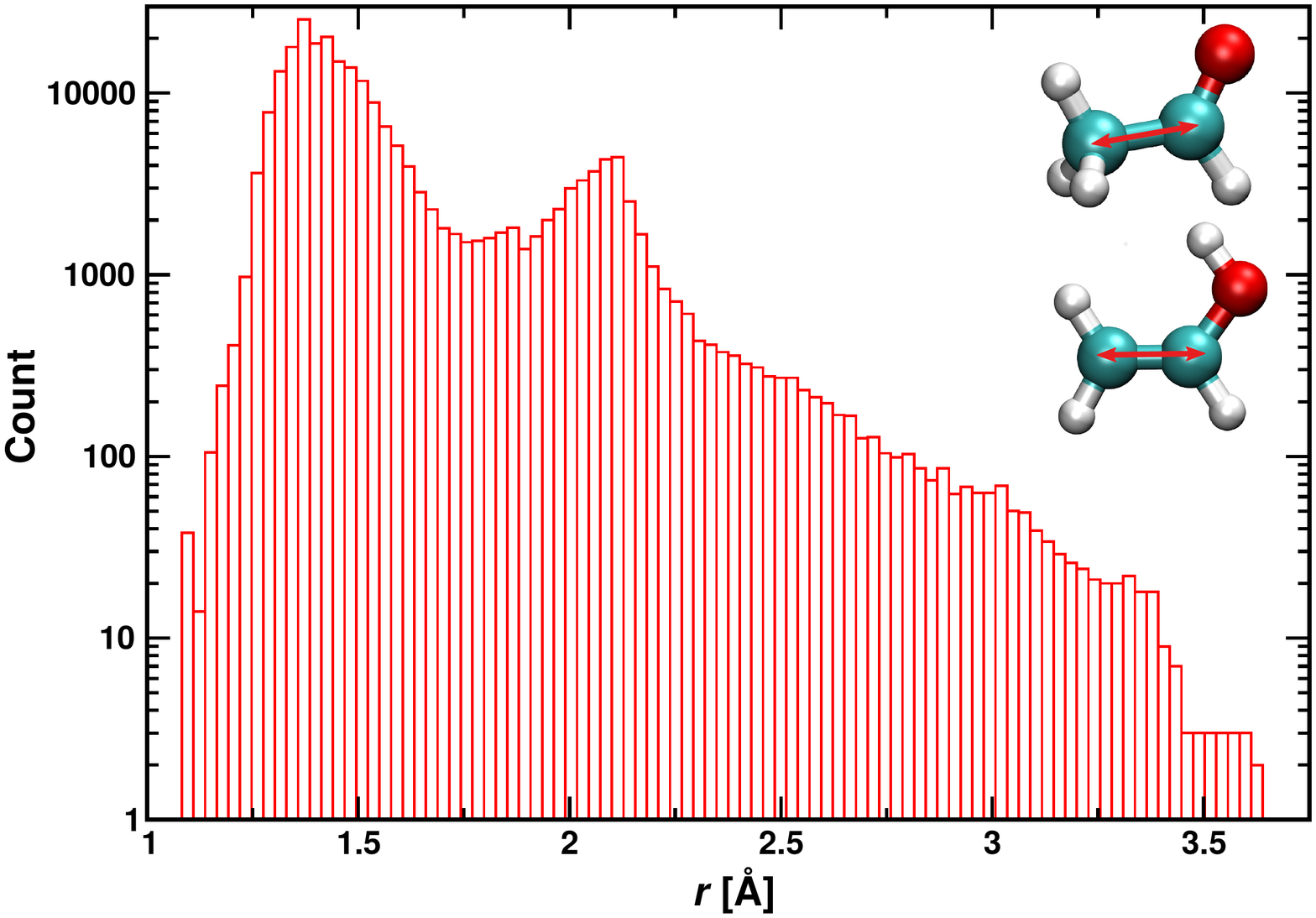}
\caption{Histogram of the CC bond lengths (red arrows) for molecules
  VA (bottom) and AA (top) molecules.}
\label{sifig:cc}
\end{figure}

\begin{figure}[ht]
\centering
\includegraphics[width=0.7\textwidth]{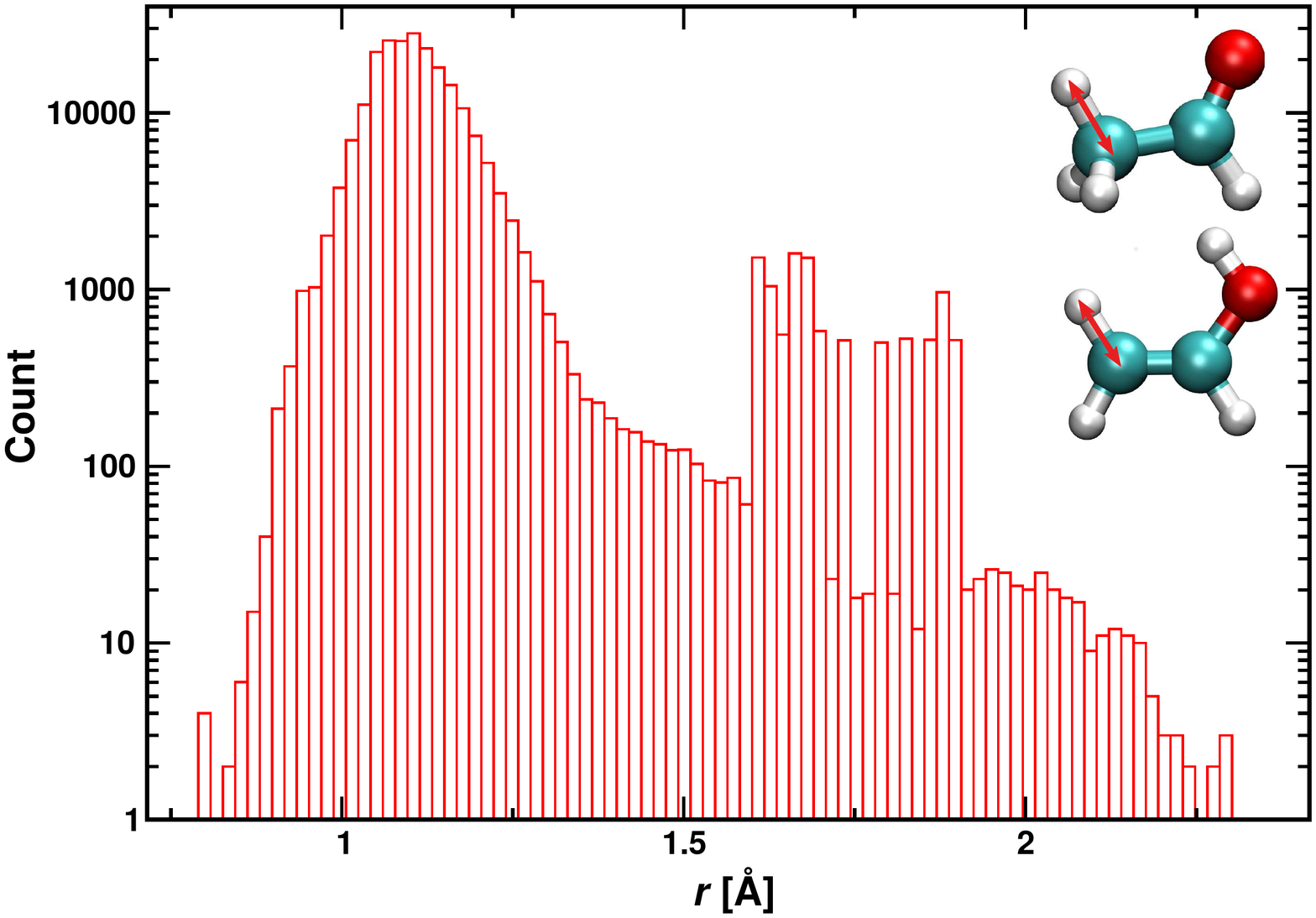}
\caption{Histogram of one methyl/methylene CH bond (red arrows)
  lengths for molecules VA (bottom) and AA (top) molecules.}
\label{sifig:ch}
\end{figure}

\begin{figure}[ht]
\centering
\includegraphics[width=0.7\textwidth]{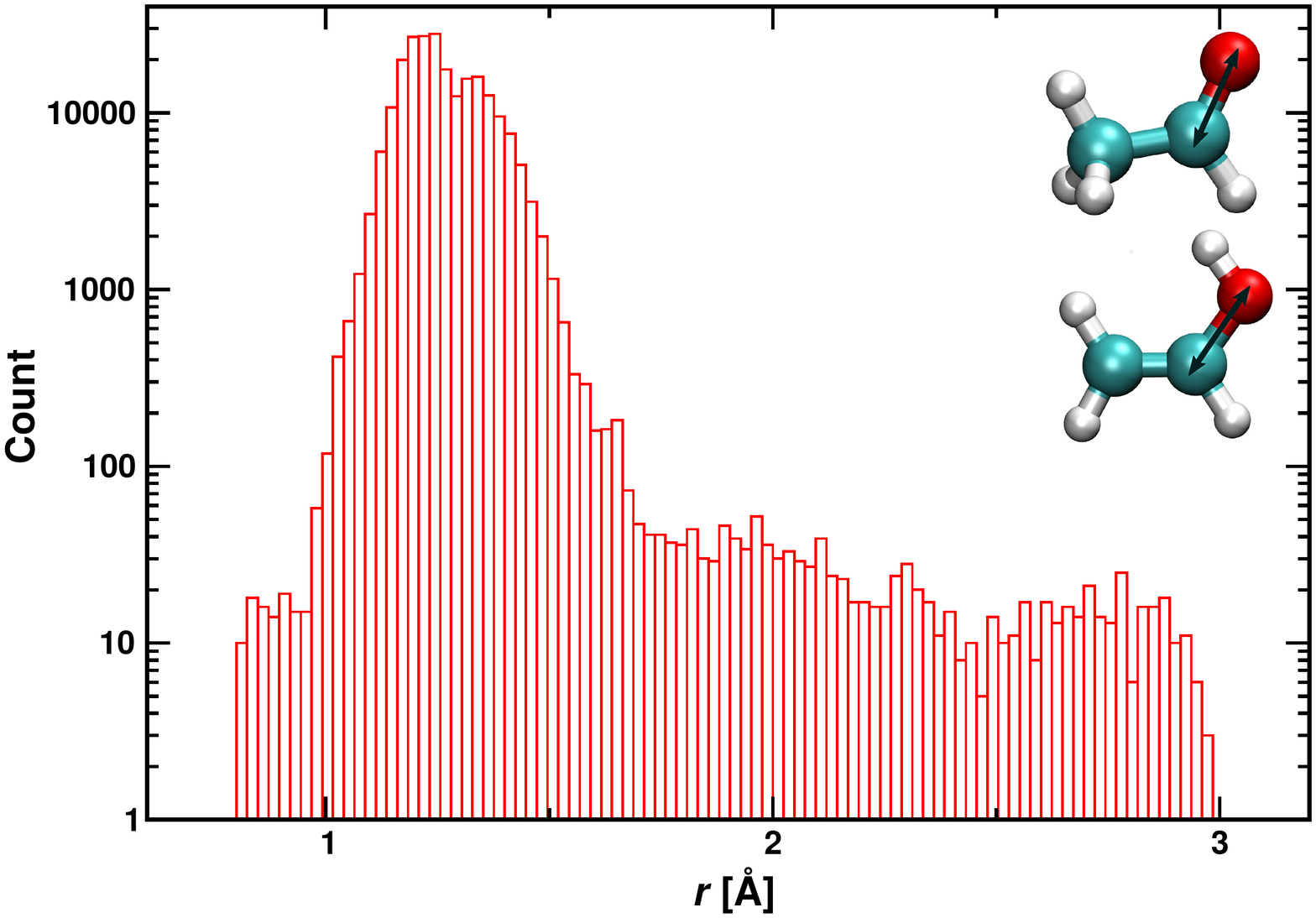}
\caption{Histogram of the CO bond lengths (black arrows) for molecules
  VA (bottom) and AA (top) molecules.}
\label{sifig:co}
\end{figure}

\begin{figure}[ht]
\centering
\includegraphics[width=0.7\textwidth]{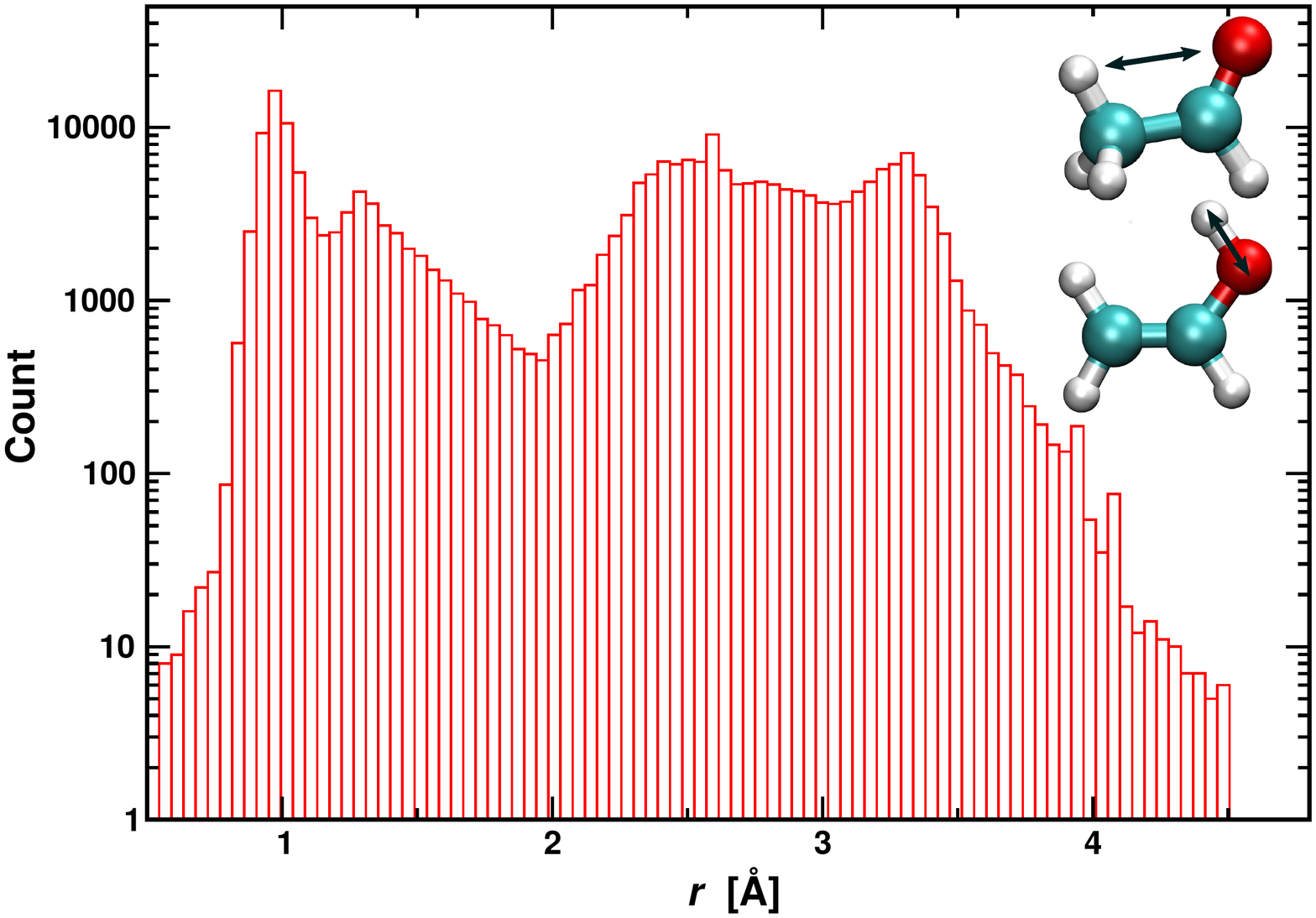}
\caption{Histogram of the OH bond lengths (black arrows) for molecules
  VA (bottom) and AA (top) molecules.}
\label{sifig:oh}
\end{figure}

\begin{figure}[ht]
\centering
\includegraphics[width=0.7\textwidth]{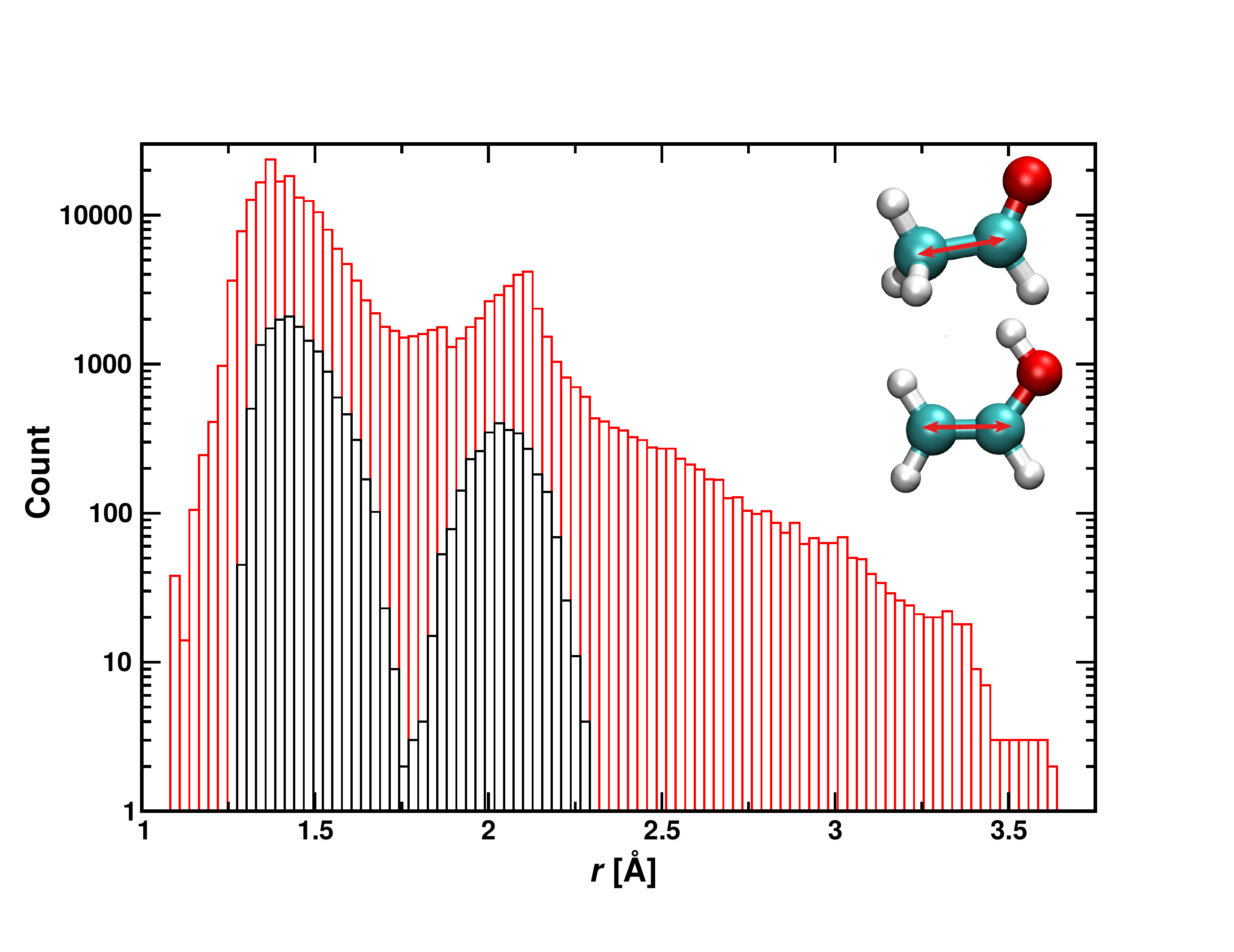}
\caption{Histogram of the CC bond lengths (red arrows) for molecules
  VA (bottom) and AA (top) molecules contained in the old data set
  with 411'204 structures. The black histogram illustrates the
  difference to the new data set.}
\label{sifig:cc_old}
\end{figure}

\begin{figure}[ht]
\centering
\includegraphics[width=0.7\textwidth]{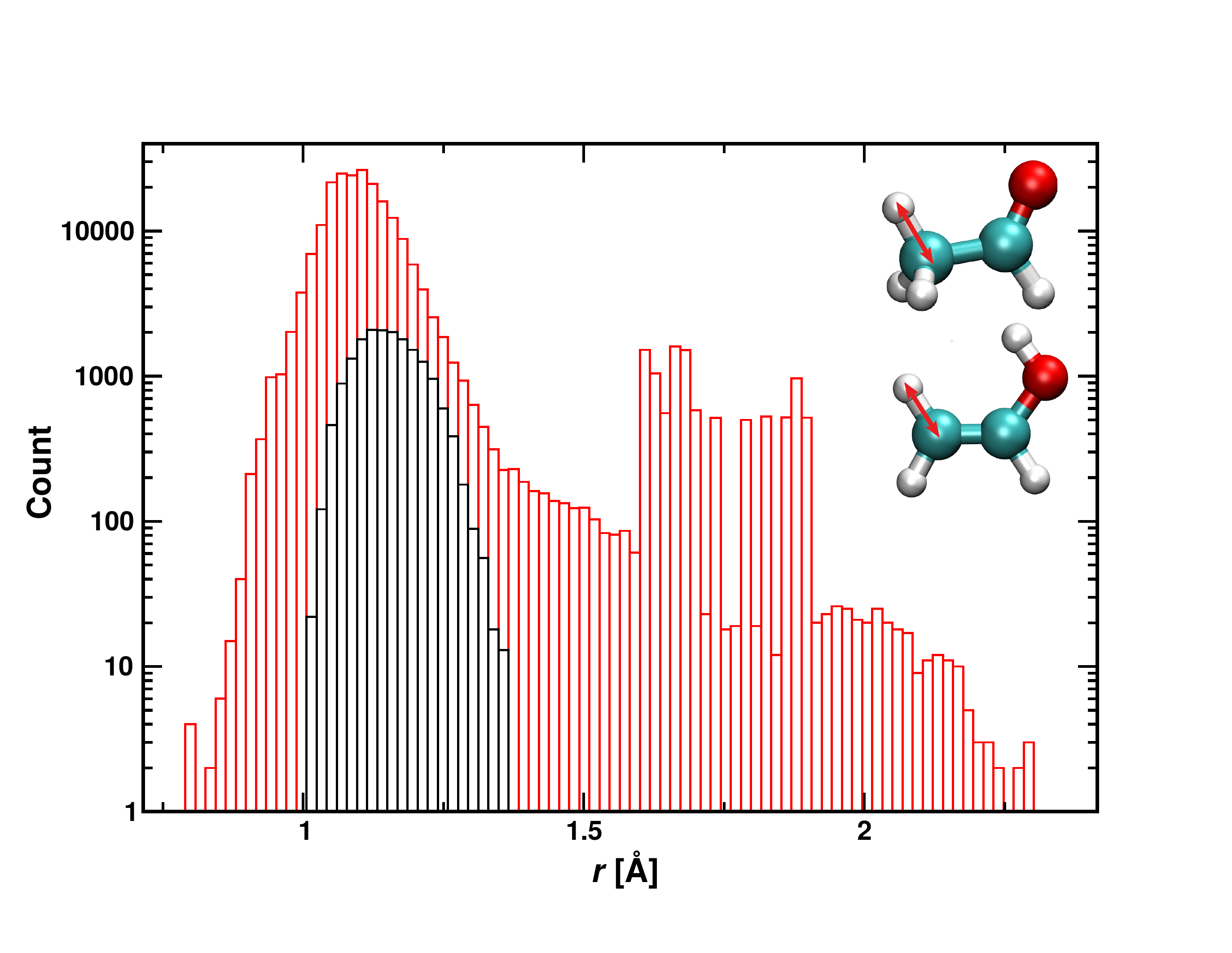}
\caption{Histogram of one methyl/methylene CH bond (red arrows)
  lengths for molecules VA (bottom) and AA (top) molecules contained
  in the old data set with 411'204 structures. The black histogram
  illustrates the difference to the new data set.}
\label{sifig:ch_old}
\end{figure}

\begin{figure}[ht]
\centering
\includegraphics[width=0.7\textwidth]{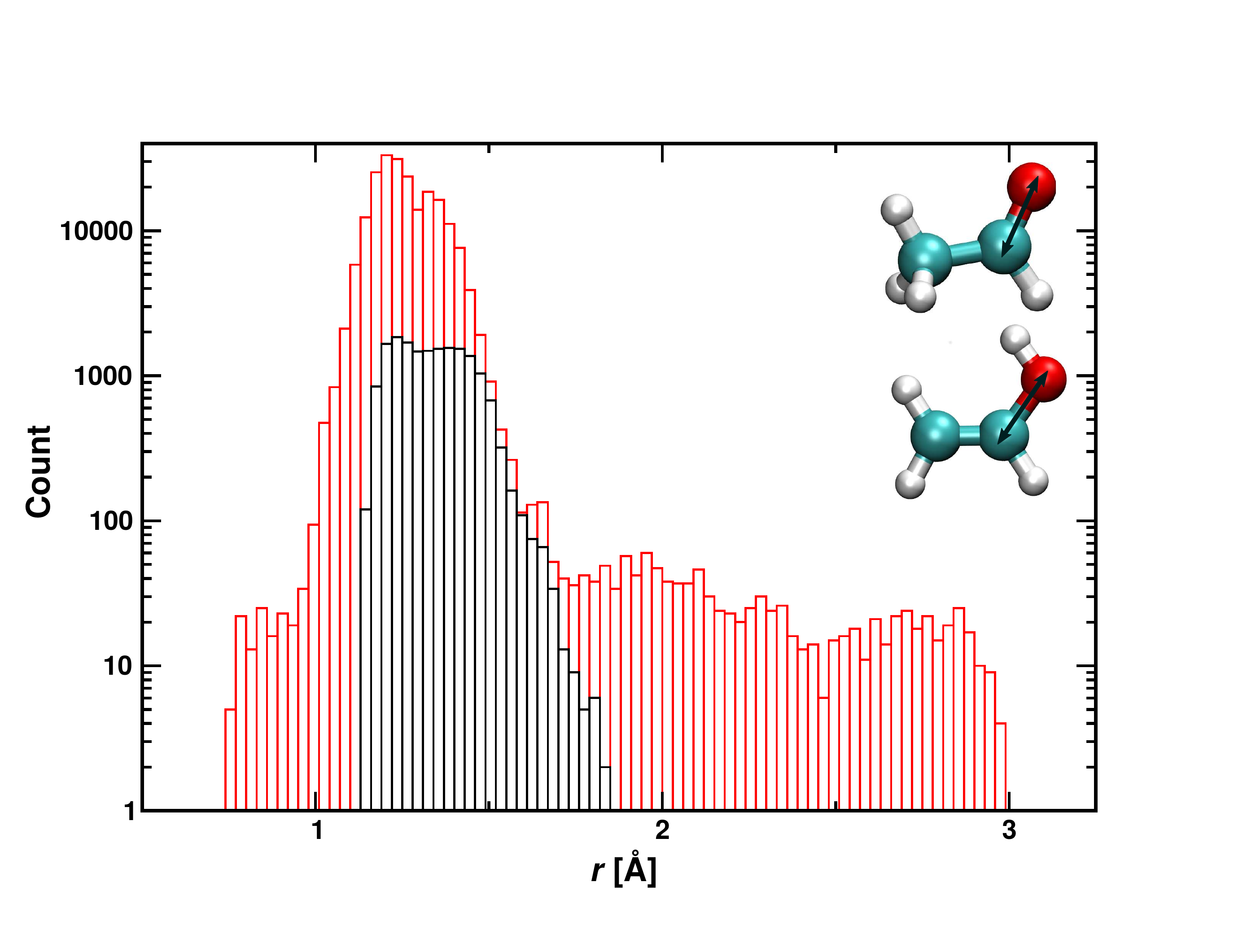}
\caption{Histogram of the CO bond lengths (black arrows) for molecules
  VA (bottom) and AA (top) molecules contained in the old data set
  with 411'204 structures. The black histogram illustrates the
  difference to the new data set.}
\label{sifig:co_old}
\end{figure}

\begin{figure}[ht]
\centering
\includegraphics[width=0.7\textwidth]{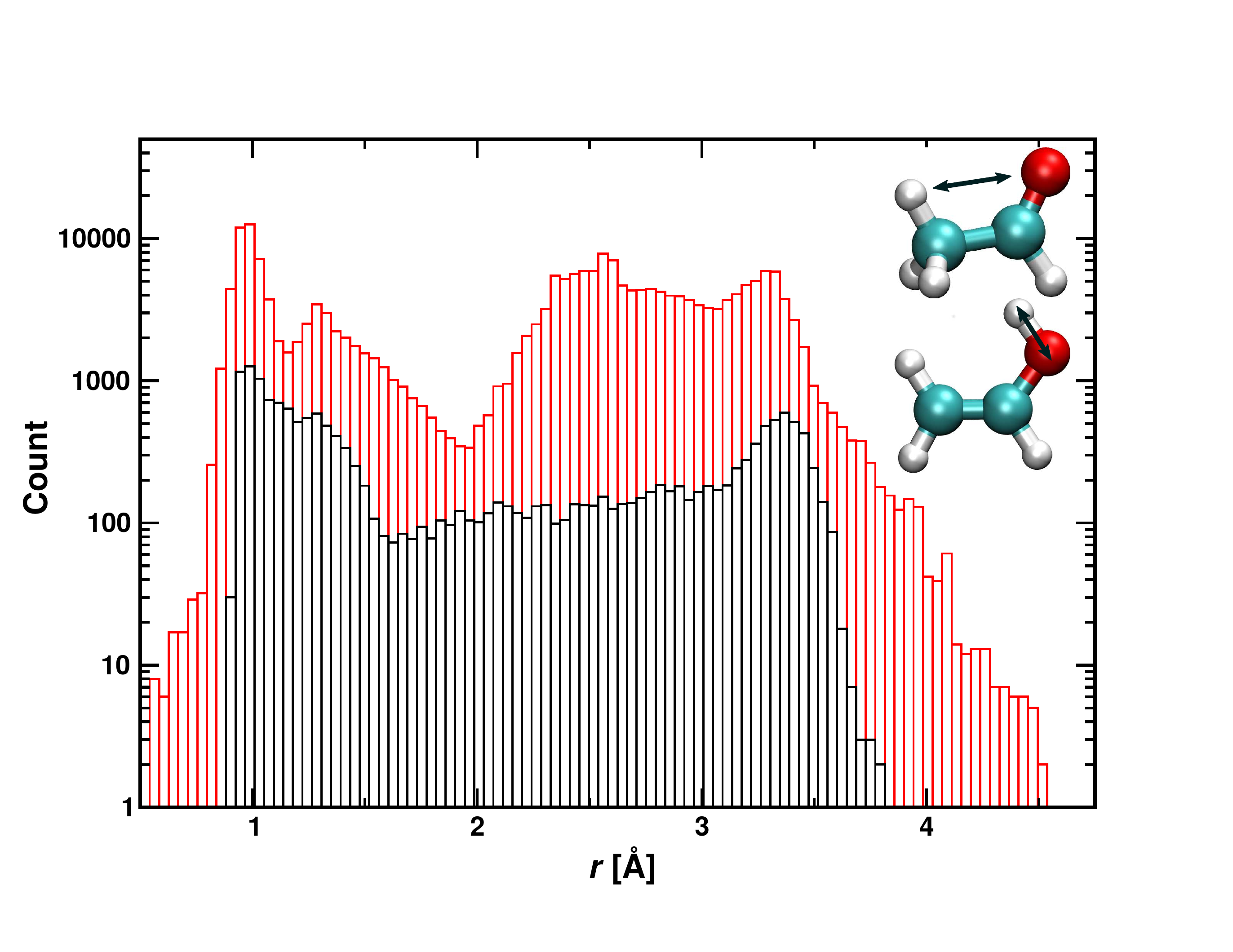}
\caption{Histogram of the OH bond lengths (black arrows) for molecules
  VA (bottom) and AA (top) molecules contained in the old data set
  with 411'204 structures. The black histogram illustrates the
  difference to the new data set.}
\label{sifig:oh_old}
\end{figure}

\clearpage
\section{RRKM Parameters}
\begin{table}[h]
\begin{tabular}{c|c|c}
Molecule             & \textit{E} [kcal/mol] & \textit{E}$_{\rm ZPE}$ [kcal/mol]\\\hline
VA           & 10.1   &  35.6  \\
AA           & 0.0    & 35.1    \\
CH$_4$+CO    & -2.2   &    \\
ketene+H$_2$ & 33.8   &    \\
TS1          & 68.1   & 34.8   \\
TS2          & 88.2   & 31.8   \\
TS3          & 84.2   & 33.2  
\end{tabular}
\caption{Energies for the optimum and TS structures used in the RRKM
  calculation based on the NN trained at the MP2/aug-cc-pVTZ level of
  theory. For the RRKM calculation, the energy barrier is ZPE
  corrected.}
\label{sitab:rrkm_energies}
\end{table}

\begin{table}[h]
\begin{tabular}{c|c|c|c|c}
VA	&	AA	&	TS1	&	TS2	&	TS3	\\\hline
445.8	&	161.4	&	2202.6i	&	1731.0i	&	1514.0i	\\
484.4	&	506.7	&	595.3	&	132.5	&	484.8	\\
714.8	&	779.8	&	637.3	&	312.1	&	571.0	\\
823.5	&	905.4	&	783.8	&	489.8	&	684.5	\\
958.1	&	1136.3	&	971.4	&	551.9	&	866.2	\\
996.2	&	1143.8	&	1062.8	&	740.4	&	960.3	\\
1118.4	&	1390.7	&	1149.4	&	886.8	&	1102.6	\\
1323.6	&	1426.6	&	1215.9	&	1057.8	&	1185.7	\\
1353.0	&	1481.6	&	1299.5	&	1442.3	&	1267.8	\\
1449.1	&	1493.9	&	1469.8	&	1445.0	&	1482.9	\\
1696.0	&	1763.4	&	1539.3	&	1831.4	&	1656.7	\\
3190.9	&	2954.4	&	1886.0	&	3044.3	&	1882.0	\\
3242.8	&	3069.5	&	3137.0	&	3179.0	&	2150.5	\\
3301.2	&	3148.8	&	3168.6	&	3193.6	&	3166.4	\\
3804.8	&	3200.0	&	3243.2	&	3206.9	&	3284.5	

\end{tabular}
\caption{NN-based frequencies for the optimum and TS structures used
  in the RRKM calculation.}
\label{sitab:rrkm_freq}
\end{table}
